# Extraction of In-Phase and Quadrature Components by Time-Encoding Sampling

*Y. H. Shao, S. Y. Chen, H. Z. Yang, F. Xi, H. Hong and Z. Liu*

Department of Electronic Engineering
Nanjing University of Science and Technology
Nanjing, Jiangsu 210094, PRC
(eezliu@njust.edu.cn)

## ABSTRACT

Time encoding machine (TEM) is a biologically-inspired scheme to perform signal sampling using timing. In this paper, we study its application to the sampling of bandpass signals. We propose an integrate-and-fire TEM scheme by which the in-phase ($I$) and quadrature ($Q$) components are extracted through reconstruction. We design the TEM according to the signal bandwidth and amplitude instead of upper-edge frequency and amplitude as in the case of bandlimited/lowpass signals. We show that the $I$ and $Q$ components can be perfectly reconstructed from the TEM measurements if the minimum firing rate is equal to the Landau's rate of the signal. For the reconstruction of $I$ and $Q$ components, we develop an alternating projection onto convex sets (POCS) algorithm in which two POCS algorithms are alternately iterated. For the algorithm analysis, we define a solution space of vector-valued signals and prove that the proposed reconstruction algorithm converges to the correct unique solution in the noiseless case. The proposed TEM can operate regardless of the center frequencies of the bandpass signals. This is quite different from traditional bandpass sampling, where the center frequency should be carefully allocated for Landau's rate and its variations have the negative effect on the sampling performance. In addition, the proposed TEM achieves certain reconstructed signal-to-noise-plus-distortion ratios for small firing rates in thermal noise, which is unavoidably present and will be aliased to the Nyquist band in the traditional sampling such that high sampling rates are required. We demonstrate the reconstruction performance and substantiate our claims via simulation experiments.

**Extraction of In-Phase and Quadrature Components by Time-Encoding Sampling**

*Y. H. Shao, S. Y. Chen, H. Z. Yang, F. Xi, H. Hong and Z. Liu*

## Contents

## I. INTRODUCTION

In radar, sonar and communications systems, the received signals are usually represented as

$$\begin{aligned} x(t) &= A(t)\cos\left(2\pi f_0 t + \varphi(t)\right) \\ &= x^{\mathrm{I}}(t)\cos\left(2\pi f_0 t\right) - x^{\mathrm{Q}}(t)\sin\left(2\pi f_0 t\right) \end{aligned} \tag{1}$$

where $A(t)$, $f_0$ and $\varphi(t)$ are the envelope, the center frequency and the phase of $x(t)$, respectively, and $x^{\mathrm{I}}(t) = A(t)\cos\left(\varphi(t)\right)$ and $x^{\mathrm{Q}}(t) = A(t)\sin\left(\varphi(t)\right)$ are the in-phase (*I*) and quadrature (*Q*) components of $x(t)$. For the convenience of signal processing, the bandpass signal $x(t)$ often needs to be translated into the *I* and *Q* components or the quadrature low-pass signals. This process is known as *I* and *Q* demodulation [1]. A number of techniques can be used to perform the demodulation [2], and among them, bandpass sampling is a well-known technique because it overcomes implementation difficulties in the baseband sampling-based techniques [3]. The principle of the bandpass sampling-based technique is shown in Fig.1, where the received signal is discretized directly at the radio or intermediate frequency and the digital *I* and *Q* components are produced through digitally processing the sampled signals. For $x(t)$ with the center frequency $f_0$ and the bandwidth $B_{\mathrm{BP}}$, the bandpass sampling [4] states that the required minimum sampling rate is given by $f_s = \left(2 f_0 + B_{\mathrm{BP}}\right)/I$, where $I = \lfloor f_0/B_{\mathrm{BP}} + 0.5 \rfloor$ and $\lfloor \bullet \rfloor$ denotes the floor function. With appropriate setting of $f_0$, the minimum Landau's rate $2B_{\mathrm{BP}}$ can be allocated.

Bandpass sampling, as any other schemes of amplitude-measuring sampling, operates under the control of a global clock which is at high sampling rates power consuming and in turn leads to electromagnetic interference [5, 6]. This limits its applications in the resource-limited cases [7]. Another problem is that the sampling rate $f_s$ is in general greater than $2B_{\mathrm{BP}}$. The minimum sampling rate $f_s = 2B_{\mathrm{BP}}$ can be allocated only for proper center frequency $f_0$ [4]. In addition, the sampling is highly sensitive to variations in the center frequency [8], which will have the negative effect on the sampling performance. It is also known that the thermal noise is unavoidably present in the sampling process [4, 9]. The input out-of-band noise will be aliased into the Nyquist band and affects the improvement of the reconstructed signal-to-noise ratios. The large improvement requires high sampling rate.

In the development of signal sampling theory, an alternative sampling framework, time encoding sampling, arouses wide interests recently. Different from the traditional framework by measuring the signal amplitudes at pre-defined sampling times, time encoding [10]-[23] makes sampling by recording the time at which the signal or its functional takes on a preset value. One of the main advantages of the sampling framework is its asynchronous operation mode, *i.e.*, no global clock is required. This advantage leads to lower power consumption and reduced electromagnetic interference.

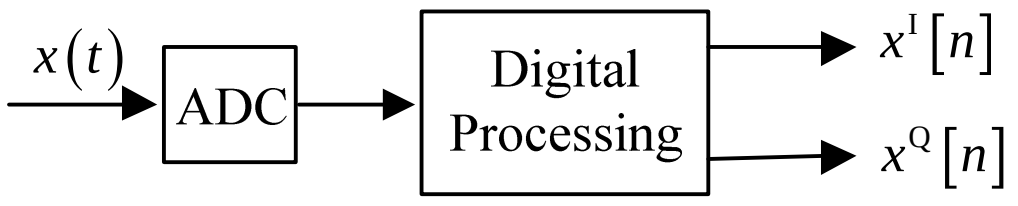

Fig.1. Bandpass Sampling-based Structure

Time encoding sampling falls under the category of event-driven sampling [7] and its realization is termed as time encoding machine (TEM) [24]. TEMs have different model structures, such as zero-crossing detectors, delta-modulation schemes and neuron-model based paradigms. Among them, TEM with integrate-and-fire (IAF) model [10] is of particular interest because it mimics the IAF mechanism of neurons in the human brain and constructs a biological neuron-like scheme to perform signal sampling. For brevity, TEM in the following refers to the TEM with integrate-and-fire (IAF) model.

TEM has a simple structure described by three parameters: bias, scale and threshold [10,17]. In operation, TEM adds a bias to its input, scales the sum and integrates the result, then compares the integral value with a threshold. A threshold-crossing or firing time is recorded when the integral reaches the threshold. Then TEM outputs a firing sequence. It is important that the sequence contains the information enough to reconstruct the signal under appropriate conditions. It is found that for the bandlimited signal, it can be perfectly reconstructed if the minimum firing rate (the firing rate refers to the number of firing times per unit time) is larger than or equal to the Nyquist rate of the signal [10]. This condition is closely related to the traditional Shannon-Nyquist sampling theory [25] and provides a criterion for the setting of TEM parameters.

For the signal reconstruction, it is noted that the firing sequence consists of strictly increasing times and the amplitude information of signal is encoded in the sequence. Interestingly, the sequence can be reformulated as an irregular sampling sequence with the local averages of the signal as the sequence values. Then the reconstruction algorithms [26] from non-uniformly sampling theory or irregular samples can be used to perform signal reconstruction. The pioneering frame-based algorithm [10] and the projection onto convex sets (POCS) algorithm [17,20] show great success.

Given the advantages of TEMs over conventional sampling and its complete theory for bandlimited signals [10,11,16,17,20], TEM theories are extended into signals in shift-invariant subspaces [12,13], finite-rate-of-innovation signals [18,21,23], signals in reproducing kernel subspaces [14] and even non-bandlimited signals [15]. In this paper, we study TEM theory for bandpass signals. Even though the signals can be sampled by the bandlimited TEM scheme, the TEM specific to the bandpass signals would have some advantages as bandpass sampling. We propose a sampling setup shown in Fig.2, which is similar to Fig.1 in some sense and is termed as Bandpass TEM (BP-TEM) for convenience. In the proposed scheme, the received signal is sampled by TEM at the radio or intermediate frequency and the reconstruction algorithm is used to reconstruct the $I$ and $Q$ components. Different from TEMs for bandlimited signals, BP-TEM parameters are set according to the signal bandwidth and amplitude instead of the upper-edge frequency and amplitude. We study the samplability of bandpass signals by BP-TEM, and show that the $I$ and $Q$ components can be perfectly reconstructed from the TEM outputs if the minimum firing rate is equal to the Landau's rate of the bandpass sampling [27]. To reconstruct the $I$ and $Q$ components, we develop an alternating projection onto convex sets (POCS) algorithm in which two POCS algorithms are alternately iterated. For the convergence analysis, we define a solution space of vector-valued signals for the projector formulation and prove that the proposed reconstruction algorithm converges to the correct unique solution if the firing rate is larger than or equal to the Landau's rate.

In comparison with traditional bandpass sampling, the BP-TEM has two typical advantages in addition to inherent properties of lower power consumption and reduced electromagnetic interference. The first one is that the BP-TEM operates regardless of the center frequencies of the bandpass signals. This is quite different from traditional bandpass sampling, where the center frequency should be carefully allocated for Landau's rate [4] and its variations have the negative effect on the sampling performance [8]. The second is the robustness to the thermal noise. It is well-known that the thermal noise is unavoidably present in the traditional sampling [9] and will be aliased to the Nyquist band such that high sampling rates are required to keep certain signal-to-noise ratios at the sampling output. However, the BP-TEM can effectively smooth the out-of-band thermal noise and small firing rates are enough to keep the signal-to-noise ratios.

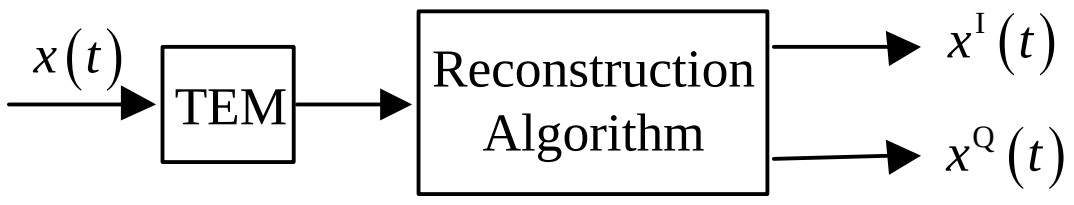

Fig.2. Bandpass Time-Encoding Machine

This paper is organized as follows. In Section II, we give the fundamentals of TEM for bandlimited signals and the POCS reconstruction algorithm. Section III devotes to the development of BP-TEM, including BP-TEM structure in Section III-A, POCS-based reconstruction of $I$ and $Q$ components in Section III-B, Convergence analysis in Section III-C and Closed-form solution in Section III-D. Extensive simulations are conducted in Section IV and conclusions are given in Section V. All simulations are reproducible using code available online [28].

*Notations*: $\mathbf{I}$ denotes the identity matrix. Boldface uppercase letters denote the matrices, and boldface lowercase letters denote the vectors. $*$ denotes the convolution operation. $(\cdot)^T$ represents the transpose operation. $\mathbb{R}$ represents the set of real values. $\mathbb{Z}$ represents the set of integer numbers.

Let $L^2(\mathbb{R})$ be the space of (scalar-valued) signals of finite energy,

$$L^2(\mathbb{R}) = \left\{ u \middle| \int_{\mathbb{R}} |u(t)|^2 \, dt < \infty \right\}$$

with norm $\|u\|^2 = \int_{\mathbb{R}} |u(t)|^2 \, dt$. Endowed with the inner product $\langle u, v \rangle = \int_{\mathbb{R}} u(t) v(t) dt$, $L^2(\mathbb{R})$ is a Hilbert space. Denote $\mathcal{B}_{\mathrm{BL}}$ and $\mathcal{B}_{\mathrm{BP}}$ as the subspaces of $L^2(\mathbb{R})$ consisting of the bandlimited signals bandlmited to $B_{\mathrm{BL}}/2$ and the bandpass signals with the center frequency $f_0$ and the bandwidth $B_{\mathrm{BP}}$, respectively.

Similarly, let $\boldsymbol{L}^2(\mathbb{R})$ be the space of (two-dimensional) vector-valued signals of finite energy,

$$\boldsymbol{L}^2(\mathbb{R}) = \left\{ \mathbf{u} \middle| \int_{\mathbb{R}} \mathbf{u}^T(t)\mathbf{u}(t)\,dt < \infty \right\}$$

with norm $\|\mathbf{u}\|^2 = \int_{\mathbb{R}} \mathbf{u}^T(t)\mathbf{u}(t)\,dt$. Then $\boldsymbol{L}^2(\mathbb{R})$ is a Hilbert space with the inner-product $\langle \mathbf{u}, \mathbf{v} \rangle = \int_{\mathbb{R}} \mathbf{u}^T(t)\mathbf{v}(t)dt$. Denote $\boldsymbol{\mathcal{B}}_{\mathrm{BL}}$ and $\boldsymbol{\mathcal{B}}_{\mathrm{BP}}$ as the subspaces of $\boldsymbol{L}^2(\mathbb{R})$ consisting of the vector-valued signals with its elements bandlmited to $B_{\mathrm{BL}}/2$ and having the center frequency $f_0$ and the bandwidth $B_{\mathrm{BP}}$, respectively.

## II. Preliminaries: TEM and POCS

In this section, we briefly introduce the TEM principle and POCS-based reconstruction algorithm for the bandlimited signal $x(t) \in \mathcal{B}_{\mathrm{BL}}$. See [10,17,20] for details.

### A. TEM Principle

The TEM principle structure is shown in Fig.3. It consists of an adder, an integrator and a comparator. Three parameters $\kappa$, $\delta$ and $b$ are used to describe the structure. For input signal $x(t) \in L^2(\mathbb{R})$, TEM adds a bias $b$ to it, scales the sum by $1/\kappa$ and then integrates the result until a threshold $\delta$. When this threshold is reached, a firing time is recorded, the value of the integrator is reset to $-\delta$. To keep regular operation, the signal $x(t)$ is assumed to be bounded, $|x(t)| \le c$, and the bias $b$ is set to be $b > c$ so that the integrator output is a positive increasing function of time. In this way, TEM outputs a firing sequence consisting of strictly increasing times $\{t_k | k \in \mathbb{Z}\}$ with $t_{k+1} > t_k$, and it satisfies

$$\frac{1}{\kappa}\int_{t_k}^{t_{k+1}} \big(x(\tau) + b\big)\,d\tau = 2\delta \tag{2}$$

Defining $y_k \triangleq \int_{t_k}^{t_{k+1}} x(\tau)\,d\tau$, we have from (2) that

$$y_k = 2\kappa\delta - b\big(t_{k+1} - t_k\big), \quad k \in \mathbb{Z} \tag{3}$$

This sequence $y_k$ can be written as

$$y_k = \langle \pi_k, x \rangle, \quad k \in \mathbb{Z} \tag{4}$$

with the indicator function $\pi_k(t)$ defined by

$$\pi_k(t) = \begin{cases} 1, & t \in [t_k, t_{k+1}) \\ 0, & \text{otherwise} \end{cases} \tag{5}$$

In this way, an amplitude-integral sequence $\{y_k | k \in \mathbb{Z}\}$ is derived. Because of non-uniform distances between any two consecutive firing times, the sequence $\{y_k | k \in \mathbb{Z}\}$ resembles irregular sampling sequence of $x(t)$. Note that $|x(t)| \le c$. Then it can be obtained from (3) that

$$\frac{2\kappa\delta}{b+c} \le t_{k+1} - t_k \le \frac{2\kappa\delta}{b-c} \tag{6}$$

That is, the difference between any two consecutive values of the sequence $\{t_k | k \in \mathbb{Z}\}$ is bounded between $2\kappa\delta/(b+c)$ and $2\kappa\delta/(b-c)$. Then we have the firing rate $F_R$ satisfying

$$\frac{b-c}{2\kappa\delta} \le F_R \le \frac{b+c}{2\kappa\delta} \tag{7}$$

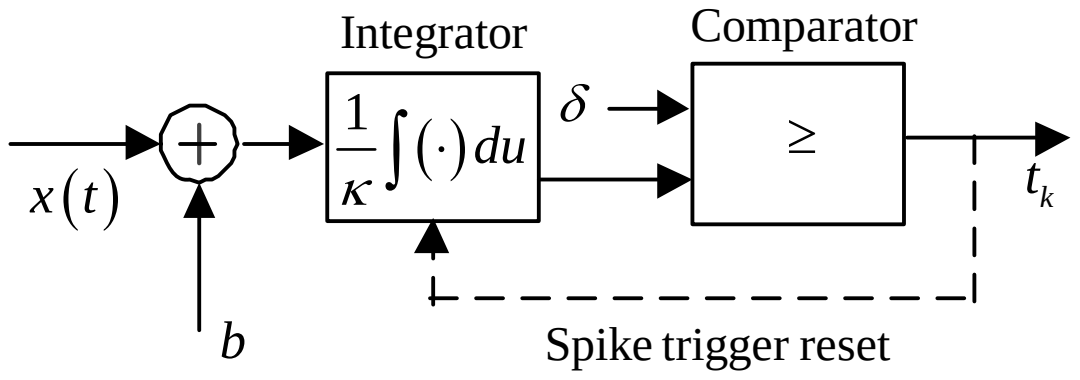

Fig.3. Time encoding machine with spike trigger reset

A fundamental problem in TEM theory is if the sequence $\{(t_k, y_k) | k \in \mathbb{Z}\}$ contains the information enough to reconstruct the signal $x(t)$ or under what conditions the signal can be reconstructed from the sequence. For bandlimited signals $x(t) \in \mathcal{B}_{\text{BL}}$, let us denote $g_{\text{BL}}(t) = \sin(\pi B_{\text{BL}} t)/(\pi t)$ as the impulse response of an ideal low-pass filter with the bandwidth $B_{\text{BL}}/2$. Then the function $f_{\text{BL}k}(t) = g_{\text{BL}}(t) * \pi_k(t)$ is the orthogonal projection of $\pi_k(t)$ onto $\mathcal{B}_{\text{BL}}$. Consequently, $\langle \pi_k - f_{\text{BL}k}, u \rangle = 0$ for all $u \in \mathcal{B}_{\text{BL}}$ and

$$\langle \pi_k, u \rangle = \langle f_{\text{BL}k}, u \rangle, \quad u \in \mathcal{B}_{\text{BL}} \tag{8}$$

The signal set defined by (9) will include all possible bandlimited signals that generate $\{(t_k, y_k) | k \in \mathbb{Z}\}$,

$$\mathcal{S}_{\text{BL}} = \{ u \in \mathcal{B}_{\text{BL}} | \langle f_{\text{BL}k}, u \rangle = y_k, k \in \mathbb{Z} \} \tag{9}$$

Thus $x(t) \in \mathcal{S}_{\text{BL}}$. If the set $\mathcal{S}_{\text{BL}}$ contains a single element, it can be expected that the signal $x(t)$ can be uniquely determined from the sequence $\{(t_k, y_k) | k \in \mathbb{Z}\}$. This was proved in [20] under the condition

$$B_{\text{BL}} \le \frac{b-c}{2\kappa\delta} \tag{10}$$

Therefore, the bandlimited signal $x(t) \in \mathcal{B}_{\text{BL}}$ is recoverable if the TEM parameters are set such that the inverse of the maximum difference between any two consecutive values of the sequence $\{t_k | k \in \mathbb{Z}\}$ is larger than or equal to $B_{\mathrm{BL}}$.

### B. POCS Reconstruction Algorithm

The reconstruction of $x(t)$ is to find a feasible point in the solution set $\mathcal{S}_{\mathrm{BL}}$. We now consider a POCS-based approach [17, 20] which searches for a fixed point in the intersection of convex sets by alternately projecting on each set. Therefore, it is fundamental to formulate the solution set as the intersection of convex sets and to identify the corresponding projection operators on the convex sets.

For the decomposition of solution space, it is noted that (9) can be formulated as

$$\mathcal{S}_{\mathrm{BL}} = \mathcal{A} \cap \mathcal{B}_{\mathrm{BL}} \tag{11}$$

where

$$\mathcal{A} = \left\{ u \in L^2(\mathbb{R}) \middle| \langle \pi_k, u \rangle = y_k, \forall k \in \mathbb{Z} \right\} \tag{12}$$

Then $x(t)$ lies in the intersection of $\mathcal{A}$ and $\mathcal{B}_{\mathrm{BL}}$. It is shown that both $\mathcal{A}$ and $\mathcal{B}_{\mathrm{BL}}$ are convex [17].

Denote $\mathcal{R}_{\mathcal{A}}$ and $\mathcal{R}_{\mathcal{B}_{\mathrm{BL}}}$ as the projection operators on $\mathcal{A}$ and $\mathcal{B}_{\mathrm{BL}}$, respectively. Then by the POCS principle, we can find POCS solution for $x(t)$ iteratively,

$$x_{\ell+1}(t) = \mathcal{R}_{\mathcal{B}_{\mathrm{BL}}} \mathcal{R}_{\mathcal{A}}\left(x_\ell(t)\right), \quad x_0(t) = \mathcal{R}_{\mathcal{B}_{\mathrm{BL}}} \mathcal{R}_{\mathcal{A}}\left(x(t)\right) \tag{13}$$

if $\mathcal{R}_{\mathcal{A}}$ and $\mathcal{R}_{\mathcal{B}_{\mathrm{BL}}}$ are firmly nonexpansive [29,30]. To identify $\mathcal{R}_{\mathcal{A}}$ and $\mathcal{R}_{\mathcal{B}_{\mathrm{BL}}}$, it is noted that the two operators have different roles in the iterative process. For the operator $\mathcal{R}_{\mathcal{B}_{\mathrm{BL}}}$, its function is to project any signal $u \in L^2(\mathbb{R})$ on the bandlimited signal set $\mathcal{B}_{\mathrm{BL}}$. Then it is natural to define

$$\mathcal{R}_{\mathcal{B}_{\mathrm{BL}}}\left(u(t)\right) = u(t) * g_{\mathrm{BL}}(t) \tag{14}$$

For the operator $\mathcal{R}_{\mathcal{A}}$, its function is to seek any signal $u \in L^2(\mathbb{R})$ close to $x(t)$ as soon as possible. Because $\left\{(t_k, y_k) \middle| k \in \mathbb{Z}\right\}$ is the only information of $x(t)$ available, we can define a piece-wise constant approximation to $x(t)$ as

$$\Re_{\mathrm{BL}}\left(x(t)\right) = \sum_{k\in\mathbb{Z}} \int_{t_k}^{t_{k+1}} x(\tau)\, d\tau \frac{1}{t_{k+1} - t_k} \pi_k(t) = \sum_{k\in\mathbb{Z}} \frac{y_k}{t_{k+1} - t_k} \pi_k(t) \tag{15}$$

Similarly, the piece-wise constant approximation to $x(t) - u(t)$ is given by

$$\begin{aligned}\Re_{\mathrm{BL}}\left(x(t)-u(t)\right) &= \sum_{k\in\mathbb{Z}}\int_{t_k}^{t_{k+1}}\left(x(\tau)-u(\tau)\right)d\tau\frac{1}{t_{k+1}-t_k}\pi_k(t) \\ &= \sum_{k\in\mathbb{Z}}\left(y_k-\left\langle\pi_k(t),u(t)\right\rangle\right)\frac{1}{t_{k+1}-t_k}\pi_k(t)\end{aligned} \tag{16}$$

which measures in some sense the difference between $x(t)$ and $u(t)\in L^2(\mathbb{R})$. Therefore we can identify the operator $\mathcal{R}_{\mathcal{A}}$ as

$$\mathcal{R}_{\mathcal{A}}\left(u(t)\right)=u(t)+\Re_{\mathrm{BL}}\left(x(t)-u(t)\right) \tag{17}$$

It was proved that the two operators $\mathcal{R}_{\mathcal{A}}$ and $\mathcal{R}_{\mathcal{B}_{\mathrm{BL}}}$ are firmly nonexpansive [17].

With above reasoning, we can simply formalize the resulting POCS method as

$$x_{\ell+1}(t)=\left(x_\ell(t)+\Re_{\mathrm{BL}}\left(x(t)-x_\ell(t)\right)\right)*g_{\mathrm{BL}}(t),\ x_0(t)=\Re_{\mathrm{BL}}\left(x(t)\right)*g_{\mathrm{BL}}(t) \tag{18}$$

With the unique condition (10), the iterative algorithm (18) will perfectly reconstruct the input signal $x(t)$.

## III. Bandpass TEM

In this section, we discuss the proposed BP-TEM.

### A. BP-TEM and Conditions for Perfect Reconstruction

The proposed BP-TEM is shown in Fig.2, in which TEM operates in the same way as in Fig.3. For its regular operation, it is assumed that the signal $x(t)\in\mathcal{B}_{\mathrm{BP}}$ is bounded, $|x(t)|\le c$, and the bias $b$ is set to be $b>c$. The signal $x(t)$ inputted to BP-TEM is a bandpass one with the center frequency $f_0$ and the bandwidth $B_{\mathrm{BP}}$. The output from the TEM is a time sequence $\{t_k|k\in\mathbb{Z}\}$ from which a amplitude-integral sequence of $\{y_k|k\in\mathbb{Z}\}$ is generated as in (3). Different from TEM for the bandlimited signals, the BP-TEM is built at radio or intermediate frequency.

Our goal is to reconstruct the *I* and *Q* components, $x^{\mathrm{I}}(t)$ and $x^{\mathrm{Q}}(t)$, of the bandpass signal $x(t)\in\mathcal{B}_{\mathrm{BP}}$ or equivalently $x(t)\in\mathcal{B}_{\mathrm{BP}}$ from $\{t_k|k\in\mathbb{Z}\}$. To do so, it is necessary that $\{t_k|k\in\mathbb{Z}\}$ should contain enough information of $x(t)\in\mathcal{B}_{\mathrm{BP}}$. Theoretically, the guarantee can be ensured by the setting BP-TEM parameters as in Section II-A with $F_{\mathrm{NYQ}}=2\left(f_0+B/2\right)$. However, it is not practical for radio-frequency applications because this will produce small firing interval as indicated in (6).

Similarly to the case of bandlimited signals, we can define a solution set consisting of all bandpass signals generating $\{(t_k, y_k) | k \in \mathbb{Z}\}$,

$$\mathcal{S}_{\text{BP}} = \{ u \in \mathcal{B}_{\text{BP}} | \langle f_{\text{BP}k}, u \rangle = y_k, k \in \mathbb{Z} \} \tag{19}$$

where $f_{\text{BP}k}(t)$ is the orthogonal projection of $\pi_k(t)$ onto $\mathcal{B}_{\text{BP}}$. Let us denote $g_{\text{BP}}(t) = \left(\sin\left(2\pi\left(f_0 + B/2\right)t\right) - \sin\left(2\pi\left(f_0 - B/2\right)t\right)\right) / (\pi t)$ as the impulse response of an ideal bandpass filter with the center frequency $f_0$ and the bandwidth $B$. Then $f_{\text{BP}k}(t)$ can be expressed as

$$f_{\text{BP}k}(t) = g_{\text{BP}}(t) * \pi_k(t) \tag{20}$$

Consequently, $\langle \pi_k - f_{\text{BP}k}, u \rangle = 0$ for all $u \in \mathcal{B}_{\text{BP}}$ and

$$\langle \pi_k, u \rangle = \langle f_{\text{BP}k}, u \rangle, \quad u \in \mathcal{B}_{\text{BP}} \tag{21}$$

If the space $\mathcal{S}_{\text{BP}}$ contains a single element, we can uniquely determine the signal $x(t)$ from the sequence $\{(t_k, y_k) | k \in \mathbb{Z}\}$. To derive the conditions on TEM such that $\mathcal{S}_{\text{BP}}$ has a single element, we assume that there exists another element $\hat{x}(t)$ in the space. It is clear that $x(t) \in \mathcal{S}_{\text{BP}}$, we have

$$\int_{t_k}^{t_{k+1}} \hat{x}(\tau) d\tau = \int_{t_k}^{t_{k+1}} x(\tau) d\tau \tag{22}$$

Now let us define two signals $X(t) = \int_{-\infty}^{t} x(\tau) d\tau$ and $\hat{X}(t) = \int_{-\infty}^{t} \hat{x}(\tau) d\tau$. Then we will have $x(t) = \hat{x}(t)$ if $X(t) = \hat{X}(t)$. Note that the signals $X(t)$ and $\hat{X}(t)$ are the bandpass signals having the same center frequency and bandwidth as those of $x(t)$. In addition, the sample $X(t_k)$ of $X(t)$ at $t_k$ is equal to the sample $\hat{X}(t_k)$ of $\hat{X}(t)$ because of (22). Thus, we have two bandpass signals having the samples at the firing times $\{t_k | k \in \mathbb{Z}\}$. From the classical nonuniform sampling theory, we know that $X(t_k)$ corresponds to the sample of the unique signal $X(t)$ at the time $t_k$ when $\{t_k | k \in \mathbb{Z}\}$ satisfies Landau rate $2B_{\text{BP}}$,

$$t_{k+1} - t_k \le \frac{1}{2B_{\text{BP}}} \tag{23}$$

In another word, if the firing rate satisfies

$$F_R \geq 2B_{\text{BP}} \tag{24}$$

$X(t_k)$ and $\hat{X}(t_k)$ come from the sample of the same signal at the time $t_k$. Then we have $x(t) = \hat{x}(t)$. Combing (24) with (7), we obtain

$$2B_{\text{BP}} \leq \frac{b-c}{2\kappa\delta} \tag{25}$$

by which we can perfectly reconstruct the signal $x(t) \in \mathcal{B}_{\text{BP}}$ from the sequence $\{t_k | k \in \mathbb{Z}\}$.

The recovery conditions are sufficient and are summarized as follows.

***Theorem* 1**: For a bandpass signal $x(t) \in \mathcal{B}_{\text{BP}}$ with the bandwidth $B_{\text{BP}}$ in $L^2(\mathbb{R})$, it is assumed that it is bounded such that $|x(t)| \leq c$ and it has a well-defined integral $\int_{-\infty}^{t} x(\tau)d\tau < \infty$. Let $x(t)$ pass through a BP-TEM with parameters $\kappa$, $\delta$ and $b$, and denote the BP-TEM output series of times as $\{t_k | k \in \mathbb{Z}\}$. Then if the BP-TEM parameters are set such that $b > c$ and $2B_{\text{BP}} \leq (b-c)/(2\kappa\delta)$, the signal $x(t)$ can be uniquely recovered from the series $\{t_k | k \in \mathbb{Z}\}$ regardless of the center frequency $f_0$ of $x(t)$.

Note that the conditions in theorem 1 have nothing to do with the center frequency $f_0$ and the Landau rate is achievable. This is different from traditional bandpass sampling [4, 25] by which the Landau rate is set only under proper relations between $f_0$ and $B_{\text{BP}}$.

### B. POCS-Based Extraction of *I* and *Q* Components

We now develop a POCS-based approach for direct extraction of $I$ and $Q$ components from $\{(t_k, y_k) | k \in \mathbb{Z}\}$. As in Subsection II-B, we should identify the convex sets and the corresponding projection operators for the extraction.

Intuitively, the bandpass signal $x(t)$ can be approximated from TEM outputs in the form of piece-wise constants as in (15)

$$x_0(t) = \sum_{k \in \mathbb{Z}} y_k \frac{\pi_k(t)}{t_{k+1} - t_k} \tag{26}$$

We can then demodulate $x_0(t)$ or $\pi_k(t)$ and obtain the approximate $I$ and $Q$ components from (26). With this observation, let us define two functions $h_k^{\text{I}}(t) = \pi_k(t)\cos(2\pi f_0 t)$ and $h_k^{\text{Q}}(t) = \pi_k(t)\sin(2\pi f_0 t)$. Then, the signals $f_k^{\mathrm{I}}(t) = h_k^{\mathrm{I}}(t) * g_{\mathrm{BL}}(t)$ and $f_k^{\mathrm{Q}}(t) = h_k^{\mathrm{Q}}(t) * g_{\mathrm{BL}}(t)$ corresponds to the *I* and *Q* components of $\pi_k(t)$, respectively. In fact, the functions $f_k^{\mathrm{I}}(t)$ and $f_k^{\mathrm{Q}}(t)$ are the orthogonal projections of $h_k^{\mathrm{I}}(t)$ and $h_k^{\mathrm{Q}}(t)$ onto $\mathcal{B}_{\mathrm{BL}}$, respectively. Consequently, $\langle h_k^{\mathrm{I}} - f_k^{\mathrm{I}}, u^{\mathrm{I}} \rangle = 0$ for all $u^{\mathrm{I}} \in \mathcal{B}_{\mathrm{BL}}$ and $\langle h_k^{\mathrm{Q}} - f_k^{\mathrm{Q}}, u^{\mathrm{Q}} \rangle = 0$ for all $u^{\mathrm{Q}} \in \mathcal{B}_{\mathrm{BL}}$. Then the solution set (19) can be equivalently described by the space,

$$\mathcal{S}^{\mathrm{I,Q}} = \left\{ \left(u^{\mathrm{I}}(t), u^{\mathrm{Q}}(t)\right) \middle| \left\langle f_k^{\mathrm{I}}, u^{\mathrm{I}} \right\rangle - \left\langle f_k^{\mathrm{Q}}, u^{\mathrm{Q}} \right\rangle = y_k, k \in \mathbb{Z}, u^{\mathrm{I}}(t) \in \mathcal{B}_{\mathrm{BL}}, u^{\mathrm{Q}}(t) \in \mathcal{B}_{\mathrm{BL}} \right\} \quad (27)$$

By (27), we can formulate the solution spaces for *I* and *Q* components respectively as

$$\mathcal{S}^{\mathrm{I}} = \left\{ u^{\mathrm{I}}(t) \in \mathcal{B}_{\mathrm{BL}} \middle| \left\langle f_k^{\mathrm{I}}, u^{\mathrm{I}} \right\rangle - \left\langle f_k^{\mathrm{Q}}, u^{\mathrm{Q}} \right\rangle = y_k, k \in \mathbb{Z}, u^{\mathrm{Q}}(t) \in \mathcal{B}_{\mathrm{BL}} \right\} \quad (28)$$

and

$$\mathcal{S}^{\mathrm{Q}} = \left\{ u^{\mathrm{Q}}(t) \in \mathcal{B}_{\mathrm{BL}} \middle| \left\langle f_k^{\mathrm{I}}, u^{\mathrm{I}} \right\rangle - \left\langle f_k^{\mathrm{Q}}, u^{\mathrm{Q}} \right\rangle = y_k, k \in \mathbb{Z}, u^{\mathrm{I}}(t) \in \mathcal{B}_{\mathrm{BL}} \right\} \quad (29)$$

which can be further decomposed as

$$\mathcal{S}^{\mathrm{I}} = \mathcal{A}^{\mathrm{I}} \cap \mathcal{B}_{\mathrm{BL}}, \ \ \mathcal{S}^{\mathrm{Q}} = \mathcal{A}^{\mathrm{Q}} \cap \mathcal{B}_{\mathrm{BL}} \quad (30)$$

where

$$\mathcal{A}^{\mathrm{I}} = \left\{ u^{\mathrm{I}}(t) \in L^2(\mathbb{R}) \middle| \left\langle h_k^{\mathrm{I}}, u^{\mathrm{I}} \right\rangle - \left\langle h_k^{\mathrm{Q}}, u^{\mathrm{Q}} \right\rangle = y_k, k \in \mathbb{Z}, u^{\mathrm{Q}}(t) \in L^2(\mathbb{R}) \right\} \quad (31)$$

$$\mathcal{A}^{\mathrm{Q}} = \left\{ u^{\mathrm{Q}}(t) \in L^2(\mathbb{R}) \middle| \left\langle h_k^{\mathrm{I}}, u^{\mathrm{I}} \right\rangle - \left\langle h_k^{\mathrm{Q}}, u^{\mathrm{Q}} \right\rangle = y_k, k \in \mathbb{Z}, u^{\mathrm{I}}(t) \in L^2(\mathbb{R}) \right\} \quad (32)$$

Then the *I* component of the bandpass signal $x(t)$ lie in the intersection of the sets $\mathcal{A}^{\mathrm{I}}$ and $\mathcal{B}_{\mathrm{BL}}$, and the *Q* component is in the intersection of $\mathcal{A}^{\mathrm{Q}}$ and $\mathcal{B}_{\mathrm{BL}}$.

Define $\mathcal{R}_{\mathcal{A}^{\mathrm{I}}}$ as the projection operator on $\mathcal{A}^{\mathrm{I}}$. Then for a given $u^{\mathrm{Q}}(t)$ in $\mathcal{S}^{\mathrm{Q}}$, we can take POCS algorithm to find a solution for $x^{\mathrm{I}}(t)$ by the following iterations,

$$x_{\ell+1}^{\mathrm{I}}(t) = \mathcal{R}_{\mathcal{B}_{\mathrm{BL}}} \mathcal{R}_{\mathcal{A}^{\mathrm{I}}}\left(x_\ell^{\mathrm{I}}(t)\right), \ \ x_0^{\mathrm{I}}(t) = \mathcal{R}_{\mathcal{B}_{\mathrm{BL}}} \mathcal{R}_{\mathcal{A}^{\mathrm{I}}}\left(x^{\mathrm{I}}(t)\right) \quad (33)$$

where

$$\mathcal{R}_{\mathcal{A}^{\mathrm{I}}}\left(u^{\mathrm{I}}(t)\right) = u^{\mathrm{I}}(t) + \Re_{\mathrm{BL}}\left(x^{\mathrm{I}}(t) - u^{\mathrm{I}}(t)\right), \ \ u^{\mathrm{I}}(t) \in L^2(\mathbb{R}) \quad (34)$$

with

$$\begin{aligned}\Re_{\text{BL}}\left(x^{\text{I}}(t)-u^{\text{I}}(t)\right)&=\sum_{k\in\mathbb{Z}}\int_{t_k}^{t_{k+1}}\left(x^{\text{I}}(\tau)-u^{\text{I}}(\tau)\right)d\tau\frac{1}{t_{k+1}-t_k}\pi_k(t)\\&=\sum_{k\in\mathbb{Z}}\left(\left\langle\pi_k(t),x^{\text{I}}(t)\right\rangle-\left\langle\pi_k(t),u^{\text{I}}(t)\right\rangle\right)\frac{1}{t_{k+1}-t_k}\pi_k(t)\end{aligned}\tag{35}$$

Similarly for a given $u^{\text{I}}(t)$ in $\mathcal{S}^{\text{I}}$, we have the iterative solution for $x^{\text{Q}}(t)$ as,

$$x^{\text{Q}}_{\ell+1}(t)=\mathcal{R}_{\mathcal{B}_{\text{BL}}}\mathcal{R}_{\mathcal{A}^{\text{Q}}}\left(x^{\text{Q}}_{\ell}(t)\right),\quad x^{\text{Q}}_{0}(t)=\mathcal{R}_{\mathcal{B}_{\text{BL}}}\mathcal{R}_{\mathcal{A}^{\text{Q}}}\left(x^{\text{Q}}(t)\right)\tag{36}$$

where $\mathcal{R}_{\mathcal{A}^{\text{Q}}}$ is the projection operator on $\mathcal{A}^{\text{Q}}$,

$$\mathcal{R}_{\mathcal{A}^{\text{Q}}}\left(u^{\text{Q}}(t)\right)=u^{\text{Q}}(t)+\Re_{\text{BL}}\left(x^{\text{Q}}(t)-u^{\text{Q}}(t)\right),\quad u^{\text{Q}}(t)\in L^2(\mathbb{R})\tag{37}$$

with

$$\begin{aligned}\Re_{\text{BL}}\left(x^{\text{Q}}(t)-u^{\text{Q}}(t)\right)&=\sum_{k\in\mathbb{Z}}\int_{t_k}^{t_{k+1}}\left(x^{\text{Q}}(\tau)-u^{\text{Q}}(\tau)\right)d\tau\frac{1}{t_{k+1}-t_k}\pi_k(t)\\&=\sum_{k\in\mathbb{Z}}\left(\left\langle\pi_k(t),x^{\text{Q}}(t)\right\rangle-\left\langle\pi_k(t),u^{\text{Q}}(t)\right\rangle\right)\frac{1}{t_{k+1}-t_k}\pi_k(t)\end{aligned}\tag{38}$$

Unfortunately, the operations (35) and (38) are not implementable because of unknown $\left\langle\pi_k(t),x^{\text{I}}(t)\right\rangle$ and $\left\langle\pi_k(t),x^{\text{Q}}(t)\right\rangle$. Note that (26) can be expressed as

$$\sum_{k\in\mathbb{Z}}y_k\frac{1}{t_{k+1}-t_k}\pi_k(t)=\Re^{\text{I}}\left(x^{\text{I}}(t)\right)\cos\left(2\pi f_0t\right)-\Re^{\text{Q}}\left(x^{\text{I}}(t)\right)\sin\left(2\pi f_0t\right)\tag{39}$$

where $\Re^{\text{I}}\left(x^{\text{I}}(t)\right)$ and $\Re^{\text{Q}}\left(x^{\text{Q}}(t)\right)$ are the approximations to $x^{\text{I}}(t)$ and $x^{\text{Q}}(t)$, respectively. We have

$$\begin{aligned}\Re^{\text{I}}\left(x^{\text{I}}(t)\right)&=\sum_{k\in\mathbb{Z}}y_k\frac{1}{t_{k+1}-t_k}\pi_k(t)\cos\left(2\pi f_0t\right)\\&=\sum_{k\in\mathbb{Z}}\int_{t_k}^{t_{k+1}}x(u)du\frac{1}{t_{k+1}-t_k}\pi_k(t)\cos\left(2\pi f_0t\right)\end{aligned}\tag{40}$$

and

$$\begin{aligned}\Re^{\text{Q}}\left(x^{\text{Q}}(t)\right)&=\sum_{k\in\mathbb{Z}}y_k\frac{1}{t_{k+1}-t_k}\pi_k(t)\left(-\sin\left(2\pi f_0t\right)\right)\\&=-\sum_{k\in\mathbb{Z}}\int_{t_k}^{t_{k+1}}x(u)du\frac{1}{t_{k+1}-t_k}\pi_k(t)\sin\left(2\pi f_0t\right)\end{aligned}\tag{41}$$

Define $\hat{x}(t)=u^{\text{I}}(t)\cos\left(2\pi f_0t\right)-u^{\text{Q}}(t)\sin\left(2\pi f_0t\right)$. Then for a given $u^{\text{Q}}(t)$ in $\mathcal{S}^{\text{Q}}$, the difference between $x^{\text{I}}(t)\in\mathcal{B}_{\text{BP}}$ and $u^{\text{I}}(t)\in\mathcal{B}_{\text{BP}}$ can be measured in some sense

by

$$\begin{aligned}\Re^{\mathrm{I}}\left(x^{\mathrm{I}}(t)-u^{\mathrm{I}}(t)\right) &= \sum_{k\in\mathbb{Z}}\int_{t_k}^{t_{k+1}}\left(x(\tau)-\hat{x}(\tau)\right)du\frac{1}{t_{k+1}-t_k}\pi_k(t)\cos(2\pi f_0 t) \\ &= \sum_{k\in\mathbb{Z}}\left(y_k-\left\langle \pi_k(t),\hat{x}(t)\right\rangle\right)\frac{1}{t_{k+1}-t_k}\pi_k(t)\cos(2\pi f_0 t)\end{aligned} \tag{42}$$

Let redefine an operator on $\mathcal{A}^{\mathrm{I}}$ as

$$\mathcal{R}_{\mathcal{A}^{\mathrm{I}}}\left(u^{\mathrm{I}}(t)\right) = u^{\mathrm{I}}(t) + \Re^{\mathrm{I}}\left(x^{\mathrm{I}}(t)-u^{\mathrm{I}}(t)\right) \tag{43}$$

which is implementable. Similarly, for a given $u^{\mathrm{I}}(t)$ in $\mathcal{S}^{\mathrm{I}}$, we can measure the difference between $x^{\mathrm{Q}}(t)\in\mathcal{B}_{\mathrm{BP}}$ and $u^{\mathrm{Q}}(t)\in\mathcal{B}_{\mathrm{BP}}$ by

$$\Re^{\mathrm{Q}}\left(x^{\mathrm{Q}}(t)-u^{\mathrm{Q}}(t)\right) = -\sum_{k\in\mathbb{Z}}\left(y_k-\left\langle \pi_k(t),\hat{x}(t)\right\rangle\right)\frac{1}{t_{k+1}-t_k}\pi_k(t)\sin(2\pi f_0 t) \tag{44}$$

and redefine an operator on $\mathcal{A}^{\mathrm{Q}}$ as

$$\mathcal{R}_{\mathcal{A}^{\mathrm{Q}}}\left(u^{\mathrm{Q}}(t)\right) = u^{\mathrm{Q}}(t) + \Re^{\mathrm{Q}}\left(x^{\mathrm{Q}}(t)-u^{\mathrm{Q}}(t)\right) \tag{45}$$

With the implementable operators $\mathcal{R}_{\mathcal{A}^{\mathrm{I}}}$ and $\mathcal{R}_{\mathcal{A}^{\mathrm{Q}}}$, we can obtain solutions to *I* and *Q* components by substituting (43) into (33) and (45) into (36), respectively. It is seen that the *I* and *Q* components are obtained by alternately iterating two POCS algorithms, one for *I* component and another for *Q* component. The algorithm is shown in Table I. For the convenience, we call it as Alternating POCS (APOCS).

| Table I: Alternating POCS Algorithm for the Extraction of *I* and *Q* Components |
|---|
| Initialize:<br>$x_0^{\mathrm{I}}(t)=\Re^{\mathrm{I}}\left(x^{\mathrm{I}}(t)\right)$<br>$x_0^{\mathrm{Q}}(t)=\Re^{\mathrm{Q}}\left(x^{\mathrm{Q}}(t)\right)$<br>$x_0(t)=x_0^{\mathrm{I}}(t)\cos(2\pi f_0 t)-x_0^{\mathrm{Q}}(t)\sin(2\pi f_0 t)$ |
| For $\ell=1,2,\cdots,$ do,<br>$x_\ell^{\mathrm{I}}(t)=\left(x_{\ell-1}^{\mathrm{I}}(t)+\Re^{\mathrm{I}}\left(x^{\mathrm{I}}(t)-x_{\ell-1}^{\mathrm{I}}(t)\right)\right)*g_{\mathrm{BL}}(t)$<br>$x_\ell^{\mathrm{Q}}(t)=\left(x_{\ell-1}^{\mathrm{Q}}(t)+\Re^{\mathrm{Q}}\left(x^{\mathrm{Q}}(t)-x_{\ell-1}^{\mathrm{Q}}(t)\right)\right)*g_{\mathrm{BL}}(t)$<br>$x_\ell(t)=x_\ell^{\mathrm{I}}(t)\cos(2\pi f_0 t)-x_\ell^{\mathrm{Q}}(t)\sin(2\pi f_0 t)$ |

As noted in Table I, if it converges, APOCS reconstructs the input bandpass signal while extracting *I* and *Q* components from $\left\{(t_k,y_k)\middle| k\in\mathbb{Z}\right\}$.

### C. Convergence Analysis

In this subsection, we follow a similar strategy to the one in [17, 29, 30] and show that the APOCS algorithm converges to a fixed point in the solution space $\mathcal{S}^{\mathrm{I,Q}}$. Then under the unique condition defined in (25), the APOCS algorithm will extract the unique correct components $x^{\mathrm{I}}(t)$ and $x^{\mathrm{Q}}(t)$.

Because of alternate iterations between $x_\ell^{\mathrm{I}}(t)$ or $x_\ell^{\mathrm{Q}}(t)$, it will be difficult to analyze the convergence of $x_\ell^{\mathrm{I}}(t)$ or $x_\ell^{\mathrm{Q}}(t)$ separately. With the notations of vector-valued signals in Section I, we now develop a vector-matrix formulation of APOCS algorithm by which the convergence of $x_\ell^{\mathrm{I}}(t)$ and $x_\ell^{\mathrm{Q}}(t)$ can be conducted jointly. To do so, we define a space $\boldsymbol{\mathcal{B}}_{\mathrm{BL}}$ consisting of bandlimited vector valued functions as

$$\boldsymbol{\mathcal{B}}_{\mathrm{BL}} = \left\{ \begin{bmatrix} u^{\mathrm{I}}(t) \\ u^{\mathrm{Q}}(t) \end{bmatrix} \middle| u^{\mathrm{I}}(t) \in \mathcal{B}_{\mathrm{BL}}, u^{\mathrm{Q}}(t) \in \mathcal{B}_{\mathrm{BL}} \right\} \tag{46}$$

and combine two signal spaces defined by (31) and (32) into one

$$\boldsymbol{\mathcal{A}} = \left\{ \begin{bmatrix} u^{\mathrm{I}}(t) \\ u^{\mathrm{Q}}(t) \end{bmatrix} \middle| \begin{matrix} \left\langle h_k^{\mathrm{I}}(t), u^{\mathrm{I}}(t) \right\rangle - \left\langle h_k^{Q}(t), u^{\mathrm{Q}}(t) \right\rangle = y_k, k \in \mathbb{Z}, \\ u^{\mathrm{I}}(t) \in L^2(\mathbb{R}), u^{\mathrm{Q}}(t) \in L^2(\mathbb{R}) \end{matrix} \right\} \tag{47}$$

Then letting

$$\mathbf{w}(t) = \left[u^{\mathrm{I}}(t), u^{\mathrm{Q}}(t)\right]^T \quad \text{and} \quad \mathbf{D}(t) = \left[\cos(2\pi f_0 t), -\sin(2\pi f_0 t)\right]^T$$

we have (43) and (45) in a vector form as

$$\boldsymbol{\mathcal{R}}_{\mathcal{A}}\left(\mathbf{w}(t)\right) = \mathbf{w}(t) + \mathfrak{R}\Big(x(t) - \left(\mathbf{D}(t)\right)^T \mathbf{w}(t)\Big) \tag{48}$$

where

$$\boldsymbol{\mathcal{R}}_{\mathcal{A}}\left(\mathbf{w}(t)\right) = \left[\mathcal{R}_{\mathcal{A}}^{\mathrm{I}}\left(u^{\mathrm{I}}(t)\right), \mathcal{R}_{\mathcal{A}}^{\mathrm{Q}}\left(u^{\mathrm{Q}}(t)\right)\right]^T \tag{49}$$

$$\mathfrak{R}\left(x(t)\right) = \begin{bmatrix} \sum_{k\in\mathbb{Z}} \int_{t_k}^{t_{k+1}} x(u)du \dfrac{1}{t_{k+1}-t_k} \pi_k(t) \cos(2\pi f_0 t), \\ -\sum_{k\in\mathbb{Z}} \int_{t_k}^{t_{k+1}} x(u)du \dfrac{1}{t_{k+1}-t_k} \pi_k(t) \sin(2\pi f_0 t) \end{bmatrix}^T \tag{50}$$

Then the iterative equations (33) and (36) are concisely described as

$$\mathbf{x}_{\ell+1}^{\mathrm{IQ}}(t) = \boldsymbol{\mathcal{R}}_{\boldsymbol{\mathcal{B}}_{\mathrm{BL}}}\left(\boldsymbol{\mathcal{R}}_{\mathcal{A}}\left(\mathbf{x}_\ell^{\mathrm{IQ}}(t)\right)\right) \tag{51}$$

where $\mathbf{x}_\ell^{\mathrm{IQ}}(t) = \left[x_\ell^{\mathrm{I}}(t), x_\ell^{\mathrm{Q}}(t)\right]^T$ and

$$\mathcal{R}_{\mathcal{B}_{\mathrm{BL}}}\left(\mathbf{w}(t)\right)=\left[u^{\mathrm{I}}(t),u^{\mathrm{Q}}(t)\right]^{T} * g_{\mathrm{BL}}(t) \tag{52}$$

We see that the convergence analysis of APOCS algorithm is to conduct the analysis on (51).

For the spaces defined by (46) and (47), together with their corresponding projectors defined by (49) and (52), we have the following lemmas.

*Lemma* 1: The projector $\mathcal{R}_{\mathcal{A}}$ is the firmly nonexpansive projection operator onto $\mathcal{A}$.

*Lemma* 2: The projector $\mathcal{R}_{\mathcal{B}_{\mathrm{BL}}}$ is the firmly nonexpansive projection operator onto $\mathcal{B}$.

*Lemma* 3: $\mathcal{A}$ is convex.

*Lemma* 4: $\mathcal{B}_{\mathrm{BL}}$ are convex.

Proofs of these lemmas are included in Appendix A.

By these lemmas, we have the following theorem.

***Theorem* 2:** Let $\mathcal{A}$ and $\mathcal{B}_{\mathrm{BL}}$ be the signal spaces, and let $\mathcal{R}_{\mathcal{A}}$ and $\mathcal{R}_{\mathcal{B}_{\mathrm{BL}}}$ be the corresponding operators. Assume that they are formulated from the signal bandwidth, the BP-TEM parameters and the BP-TEM outputs as in (47), (46), (48) and (52), respectively. Then for $b > c$, APOCS algorithm (51) or in Table I converges to a solution of the *I* and *Q* components lying in the intersection of the convex sets $\mathcal{A}$ and $\mathcal{B}_{\mathrm{BL}}$. Furthermore, for the known center frequency, APOCS algorithm obtains a solution of $x(t)$.

The Theorem 2 working with the Theorem 1 will extract the unique correct components $x^{\mathrm{I}}(t)$ and $x^{\mathrm{Q}}(t)$ of $x(t)$ from the firing sequence $\{t_k | k \in \mathbb{Z}\}$.

### D. Closed-Form Solution

We now derive a closed-form solution of $x^{\mathrm{I}}(t)$ and $x^{\mathrm{Q}}(t)$.

Define vectors $\mathbf{q} = [y_k]$, $\mathbf{g}^{\mathrm{I}}(t) = \left[g_k^{\mathrm{I}}(t)\right]$ and $\mathbf{g}^{\mathrm{Q}}(t) = \left[g_k^{\mathrm{Q}}(t)\right]$ and a matrix $\mathbf{G} = [G_{k\ell}]$ in which $y_k = \int_{t_k}^{t_{k+1}} x(\tau)\, d\tau$,

$$g_k^{\mathrm{I}}(t) = \left( \frac{\pi_k(t)}{t_{k+1} - t_k} \cos(2\pi f_0 t) \right) * g_{\mathrm{BL}}(t), \tag{53}$$

$$g_k^{\mathrm{Q}}(t) = \left( -\frac{\pi_k(t)}{t_{k+1} - t_k} \sin(2\pi f_0 t) \right) * g_{\mathrm{BL}}(t), \text{ and} \tag{54}$$

$$G_{k\ell} = \int_{t_k}^{t_{k+1}} \left( g_\ell^{\mathrm{I}}(t) \cos(2\pi f_0 t) - g_\ell^{\mathrm{Q}}(t) \sin(2\pi f_0 t) \right) dt. \tag{55}$$

We have the following.

***Theorem* 3:** Under the assumptions of Theorem 1 and Theorem 2, the *I* and *Q* components, $x^{\mathrm{I}}(t)$ and $x^{\mathrm{Q}}(t)$, of a bandpass signal $x(t)$ can be perfectly reconstructed from the firing sequence $\{t_k | k \in \mathbb{Z}\}$ as

$$x^{\mathrm{I}}(t) = \lim_{\ell \to \infty} x_\ell^{\mathrm{I}}(t) = \left( \mathbf{g}^{\mathrm{I}}(t) \right)^T \mathbf{G}^+ \mathbf{q} \tag{56}$$

$$x^{\mathrm{Q}}(t) = \lim_{\ell \to \infty} x_\ell^{\mathrm{Q}}(t) = \left( \mathbf{g}^{\mathrm{Q}}(t) \right)^T \mathbf{G}^+ \mathbf{q} \tag{57}$$

where $\mathbf{G}^+$ denotes the pseudoinverse of $\mathbf{G}$.

*Proof*: By induction, we can derive

$$x_\ell^{\mathrm{I}}(t) = \left( \mathbf{g}^{\mathrm{I}}(t) \right)^T \left( \sum_{k=0}^{\ell} (\mathbf{I} - \mathbf{G})^k \right) \mathbf{q} \tag{58}$$

$$x_\ell^{\mathrm{Q}}(t) = \left( \mathbf{g}^{\mathrm{Q}}(t) \right)^T \left( \sum_{k=0}^{\ell} (\mathbf{I} - \mathbf{G})^k \right) \mathbf{q} \tag{59}$$

Note that $\lim_{\ell \to \infty} \sum_{k=0}^{\ell} (\mathbf{I} - \mathbf{G})^k = \mathbf{G}^+$ for $\mathbf{G}$ defined in (55). Then we have (56) and (57). □

The reconstruction of the *I* and *Q* components has a simple interpretation. Let us consider $x^{\mathrm{I}}(t)$. Defining a vector $\mathbf{c} = [c_k] = \mathbf{G}^+ \mathbf{q}$ and substituting (53) into (56), we have

$$\begin{aligned} x^{\mathrm{I}}(t) &= \sum_{k \in \mathbb{Z}} c_k g_k^{\mathrm{I}}(t) \\ &= \sum_{k \in \mathbb{Z}} c_k \left( \frac{\pi_k(t)}{t_{k+1} - t_k} \cos(2\pi f_0 t) \right) * g_{\mathrm{BL}}(t) \\ &= \left( \left( \sum_{k \in \mathbb{Z}} \frac{c_k \pi_k(t)}{t_{k+1} - t_k} \right) \cos(2\pi f_0 t) \right) * g_{\mathrm{BL}}(t) \end{aligned} \tag{60}$$

Note that the term $\sum_{k \in \mathbb{Z}} c_k \pi_k(t) / (t_{k+1} - t_k)$ in last expression can be taken as an approximate bandpass signal with the center frequency $f_0$ and the bandwidth $B$. In this sense, as indicated in (60), the reconstructed *I* component $x^{\mathrm{I}}(t)$ can be taken as the output of the *I* banch demodulator with the input signal $\sum_{k\in\mathbb{Z}} c_k \pi_k(t)/(t_{k+1}-t_k)$. Similar interpretation can be given for the reconstruction of *Q* component $x^{\mathrm{Q}}(t)$.

## IV. Simulations

In this section, we conduct simulations to validate our theory and confirm its performance under different conditions. An amplitude-and-phase modulated bandpass signal is taken as an illustrative example,

$$x(t)=\frac{2\sin(2\pi f_1 t)}{\pi f_1 t}\cos\left(2\pi f_0 t+\frac{\sin(2\pi f_2 t)}{2\pi f_2 t}\right) \tag{61}$$

where $f_0$ is the carrier frequency, and $\sin(2\pi f_1 t)/(2\pi f_1 t)$ and $\sin(2\pi f_2 t)/(2\pi f_2 t)$ are the time-varying amplitude and phase, respectively. The *I* and *Q* components of $x(t)$ are given respectively by

$$x^{\mathrm{I}}(t)=\frac{2\sin(2\pi f_1 t)}{2\pi f_1 t}\cos\left(\frac{\sin(2\pi f_2 t)}{2\pi f_2 t}\right),\ x^{\mathrm{Q}}(t)=\frac{2\sin(2\pi f_1 t)}{2\pi f_1 t}\sin\left(\frac{\sin(2\pi f_2 t)}{2\pi f_2 t}\right) \tag{62}$$

The signal $x(t)$ has the maximal amplitude equal to 2 and its bandwidth $B_{\mathrm{BP}}$ is determined by the parameters $f_1$ and $f_2$.

For BP-TEM, its parameters are set according to Theorem 1. For convenience, let $\kappa=1$, $b=c+1$ with $c=2$ and then $\delta\le 1/(4B_{\mathrm{BP}})$.

We take the reconstructed signal-to-noise-plus-distortion ratio (SNDR) as performance metric,

$$\mathrm{SNDR}_{\mathrm{REC}}=\frac{\|x(t)\|_2}{\|x(t)-\bar{x}(t)\|_2}$$

where $\bar{x}(t)$ is the reconstructed signal. It is noted that $\mathrm{SNDR}_{\mathrm{REC}}$ measures the reconstruction accuracy harassed by the input noise and the reconstruction nonlinearity. When the input is noise-free, $\mathrm{SNDR}_{\mathrm{REC}}$ describes the performance of APOCS perfect reconstruction. Similarly, $\mathrm{SNDR}_{\mathrm{REC}}$ can be computed for *I* and *Q* components. To reduce the boundary effect of the reconstruction algorithm, we compute SNDR for the middle 90% of the sampling length. In Monto Carlo experiments, $\mathrm{SNDR}_{\mathrm{REC}}$ is given by averaging over 100 independent trials.

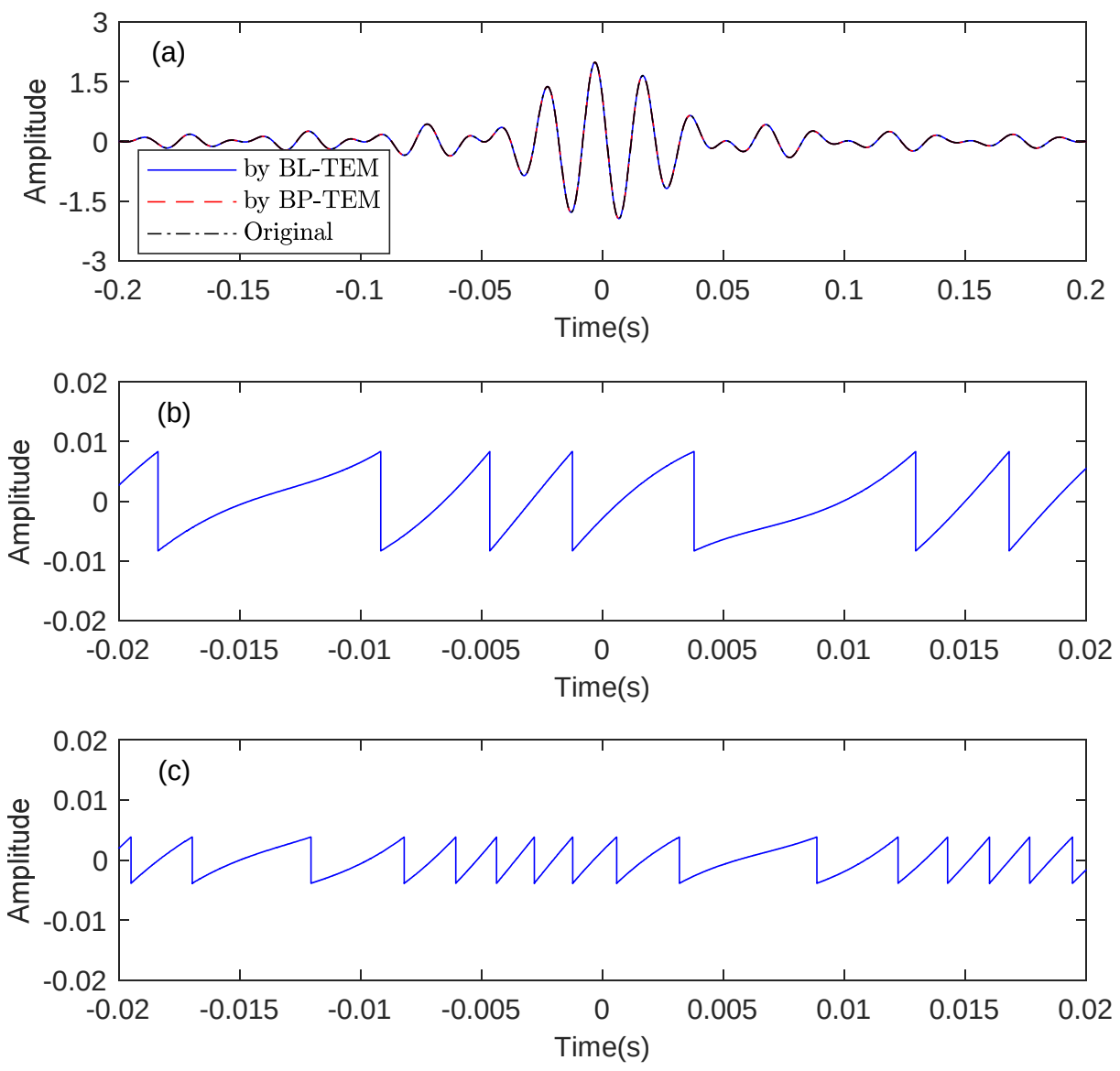

Fig. 4 Signal and its reconstruction (a) and Integrator outputs of BP-TEM (b) and BL-TEM (c)

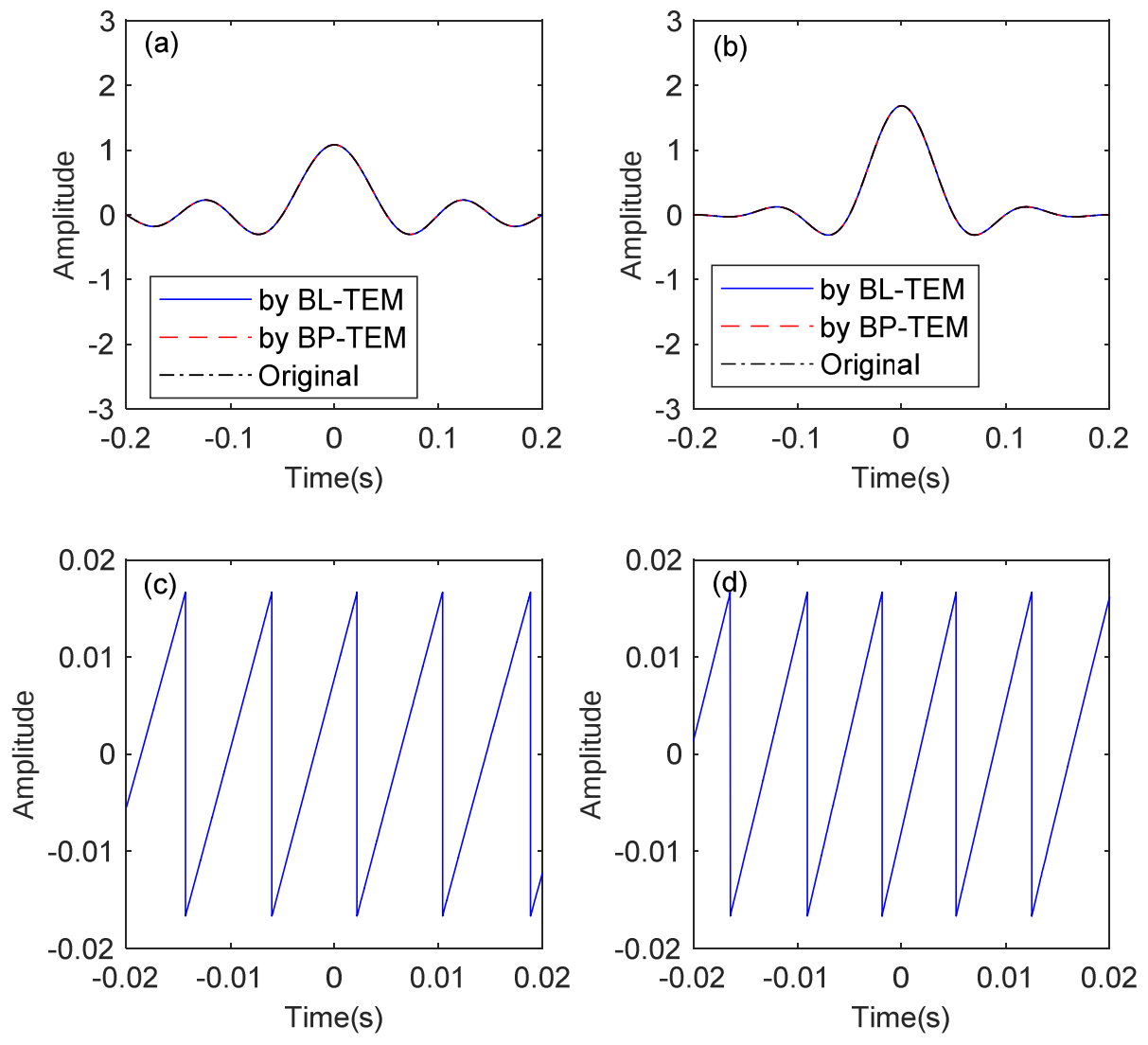

Fig. 5 *I*/*Q* components and reconstructions (a)/(b) and the Integrator outputs (c)/(d)

## A. Feasibility of TEM Sampling for Bandpass Signals

We now present the sampling and reconstruction by APOCS algorithm and make comparisons with those by POCS algorithm in [17] by considering the signal (61) as a bandlimited one. The signal parameters are set to be $f_0 = 50\text{Hz}$, $f_1 = 10\text{Hz}$ and $f_2 = 2.5\text{Hz}$, which has the approximate bandwidth of $B_{\text{BP}} = 30\text{Hz}$. The chosen $B_{\text{BP}}$ is slightly larger than -3dB bandwidth of the signal, which is convenient for the setting of TEM thresholds. For the BP-TEM in this paper and the bandlimited TEM (BL-TEM) in [17], their thresholds can be set to be $1/120$ and $1/260$, respectively. For the *I* and *Q* components by the BL-TEM, the threshold is $1/60$.

The sampling and reconstruction are shown in Fig. 4 and Fig. 5, respectively. It is seen that both BP-TEM and BL-TEM well reconstruct the signal (61) and its *I* and *Q* components (62). However, by comparing their firing sequences, as expected, the BP-TEM allows small firing rates. These results show that the proposed BP-TEM is feasible for sampling the bandpass signals.

As shown in APOCS algorithm, the reconstruction of the bandpass signal follows the reconstructions of its *I* and *Q* components. Then the reconstructed accuracy of the bandpass signal is similar to that of the *I* and *Q* components. In this sense, we will only present the reconstruction performance of the input signal to save the space.

### B. Validation of Theorem 1 and Theorem 2

To study the effect of the central frequency $f_0$ and the bandwidth $B_{\mathrm{BP}}$ on the setting of TEM parameters, we keep the parameters $f_1$ and $f_2$ intact while varying the carrier frequency $f_0$ without loss of generality. The parameters $f_1$ and $f_2$ are set as in last section and the center frequency $f_0$ is increased from $15\mathrm{Hz}$ to $1500\mathrm{Hz}$ with stepsize $1.5\mathrm{Hz}$.

The reconstruction results are shown in Fig. 6. For the upper curve, the threshold is set to be $1/120$, strictly following the signal bandwidth. It is seen that the reconstructed SNDR is kept high or the signal (61) is perfectly reconstructed with the fixed setting of BP-TEM parameters no matter what the center frequency is. That is, the Landau-rate sampling by TEM can always be achieved for any center frequency. This is quite different from classical bandpass sampling in which the center frequency and the bandwidth should be properly set for Landau-rate sampling. This property of the BP-TEM brings the ease of TEM implementation and shows its robustness to the variations in the center frequency. It is also noted that the reconstructed SNDR decreases with some fluctuations as the center frequency increases. This may be due to the implementation of BP-TEM in finite precision. The middle and lower curves give the reconstruction results for the thresholds to be $1/240$ and $1/360$. The reconstruction performance is improved for small thresholds because of more sampling numbers.

The results by Fig. 6 along with the results in Figs. 4 and 5 fully validate Theorem 1 and Theorem 2 in Section III.

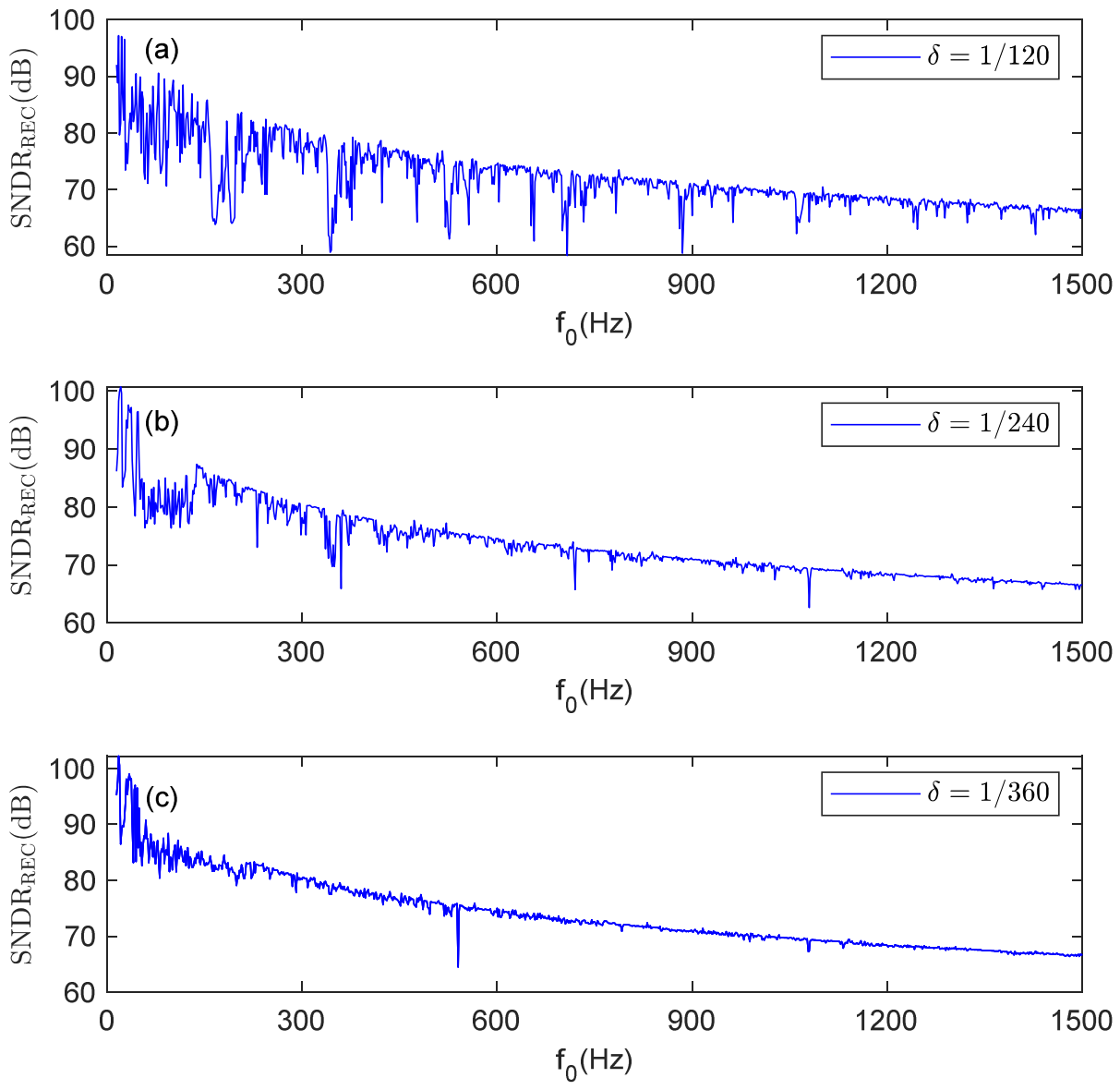

Fig. 6 Reconstructed SNDR versus center frequency: $\delta = 1/120$ (a), $\delta = 1/240$ (b) and $\delta = 1/360$ (c)

## C. Reconstruction Performance in Noise Environments

We now simulate the effect of background noise on reconstruction. Two kinds of noises are considered: white Gaussian noise and bandpass Gaussian noise. The threshold of BP-TEM to be $1/240$.

Fig. 7 (a) shows the variations of the reconstructed SNDR ($\mathrm{SNDR}_{\mathrm{REC}}$) versus the input signal-to-noise ratio ($\mathrm{SNR}_{\mathrm{IN}}$) under different center frequencies for the case of white Gaussian noise. It is seen that the reconstructed SNDR has a large increment with respect to the input SNR. This is due to the filtering effect of the operator (14). It is also noted that the center frequency has an effect on the reconstructed SNDR. In general, the smaller the center frequency is, the larger the reconstructed SNDR is. However, there are some exceptions, for example, the cases of $f_0 = 1050\mathrm{Hz}$ and $f_0 = 1500\mathrm{Hz}$. These observations agree well with the simulations in Fig. 6. The SNR increments are bounded by the in-band SNR in the input signal. For simulation example, the largest increment is 140dB for the input signal with $\mathrm{SNR}_{\mathrm{IN}} = 15\mathrm{dB}$.

In traditional bandpass sampling, the reconstructed SNDR also has an increment with respect to the input SNR, and the increments become large for high sampling rates. For example, in the simulation example, for the input signal with $f_0 = 600\text{Hz}$ and $\text{SNR}_{\text{IN}} = 15\text{dB}$, the $\text{SNDR}_{\text{REC}}$ is $15.67\text{dB}$ for the sampling rate of $65\text{Hz}$. When the sampling rate is increased to $1000\text{Hz}$, the $\text{SNDR}_{\text{REC}}$ becomes $27.76\text{dB}$. To have large increments, high sampling rates are needed. However, the BP-TEM can produce large increments even for small firing rates.

For the case of bandpass noise, its bandwidth and center frequency are set as in the input signal. Simulation results are shown in Fig. 7 (b) under the different center frequencies. It is clear that the reconstructed SNDR well keeps the input SNR. Both signal and noise get reconstructed.

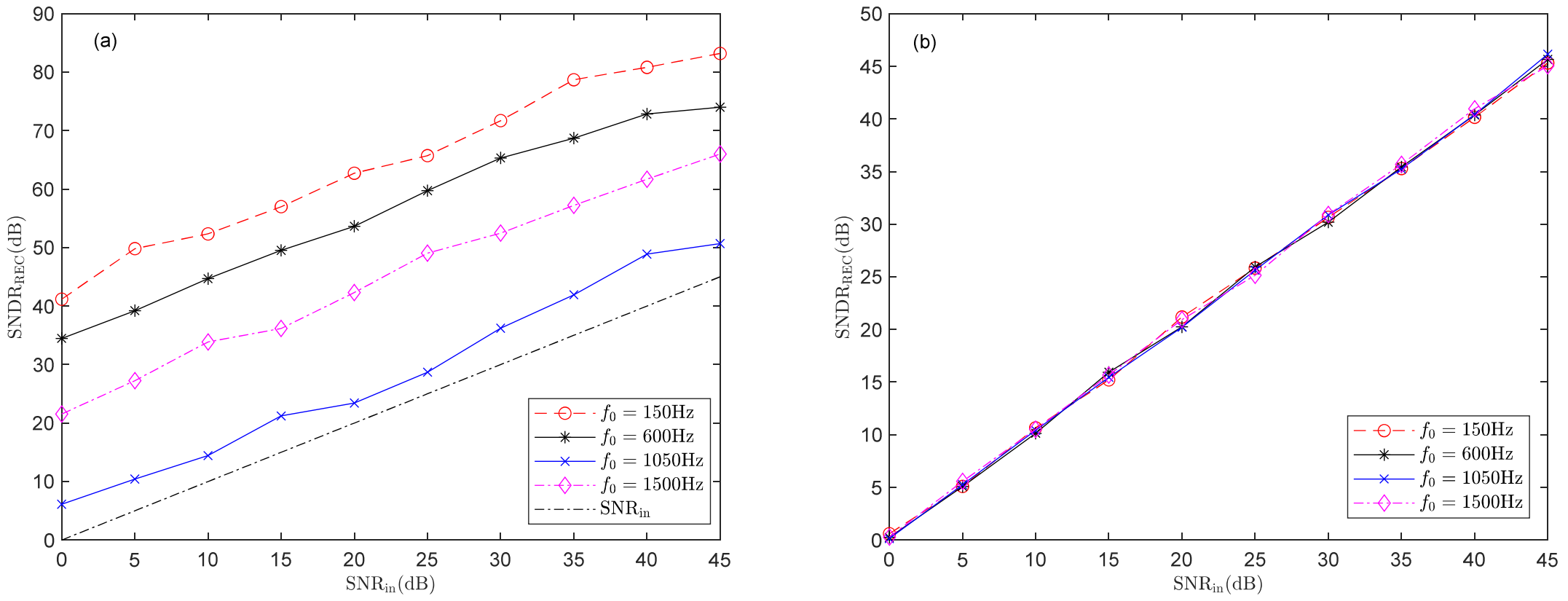

Fig. 7 Reconstructed SNDR versus input SNR

## D. Reconstruction Performance under Timing Quantization

The previous simulations assume that there are no errors in the firing sequence $\{t_k | k \in \mathbb{Z}\}$. Here we study the effect of $\{t_k | k \in \mathbb{Z}\}$ under finite precision. Following [10], the actual firing times used for signal reconstruction are modeled as $\hat{t}_k = t_k + n_k$, where $n_k$ is an independent identically distributed random variable on $[-\Delta/2, \Delta/2]$ and $\Delta$ is defined as

$$\Delta = \frac{1}{2^N}\left(\frac{2\kappa\delta}{b-c} - \frac{2\kappa\delta}{b+c}\right)$$

In the above equation, $N$ is the number of bits to represent the $T_k - T_{\min}$ with $T_k = t_{k+1} - t_k$ and $T_{\min} = 2\kappa\delta/(b+c)$. In simulations, $\delta = 1/240$ and then

$\Delta = 1/(2^N \times 150)$.

The reconstructed SNDR versus the center frequency and the number of bits is shown in Fig. 8. It may be concluded from this figure that the quantization error has small effects on low center frequencies. As the center frequencies increases, the number of bits should increase to keep the reconstruction performance.

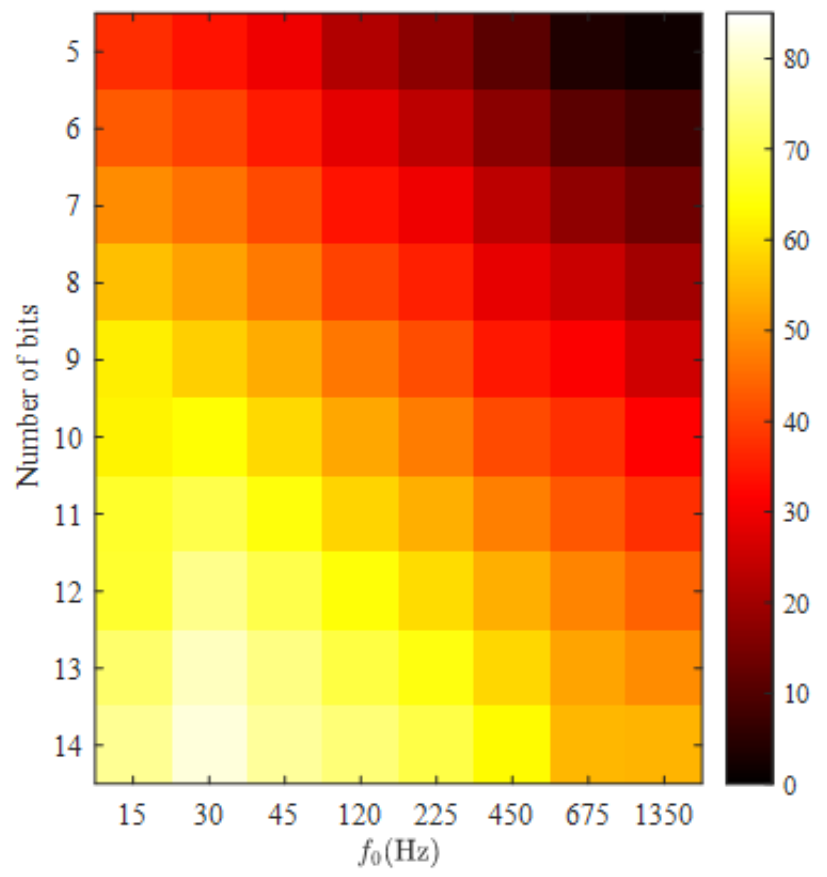

Fig. 8 Reconstructed SNDR versus center frequency and number of bits

## V. Conclusion

In this paper, we study the samplability of bandpass signals by using BP-TEMs. We first propose to design BP-TEMs according the signal bandwidth and provide sufficient conditions on signal bandwidth $B_{\mathrm{BP}}$ for the perfect reconstruction of the input bandpass signals. We then develop an APOCS algorithm to reconstruct $I$ and $Q$ components and show that it converges to the correct unique solution in the noiseless case. We finally conduct extensive simulations to validate our theory. Compared to traditional bandpass sampling for bandpass signals, our BP-TEM is robust to the variations in center frequency because it can operate regardless of center frequency. It is also robust to the thermal noise due to filtering effect in APOCS algorithm. With the inherent property of less power consuming and reduced electromagnetic interference, the proposed BP-TEM is more costly effective and has superior performance.

## Appendix A: Proofs of Lemma 1 ~ Lemma 4

*Proof of Lemma* 1: To show the lemma, it is sufficient to show that the operator is idempotent, has as the range the space $\mathcal{A}$ in $L^2(\mathbb{R})$, and is firmly nonexpansive, respectively.

First, we show that $\mathcal{R}_{\mathcal{A}}$ is idempotent. Note

$$\mathcal{R}_{\mathcal{A}}\left(\mathcal{R}_{\mathcal{A}}\left(\mathbf{w}(t)\right)\right)=\mathcal{R}_{\mathcal{A}}\left(\mathbf{w}(t)\right)+\Re\left[x(t)-\left(\mathbf{D}(t)\right)^T\mathcal{R}_{\mathcal{A}}\left(\mathbf{w}(t)\right)\right]$$

Then it is sufficient to show $\Re\left[x(t)-\left(\mathbf{D}(t)\right)^T\mathcal{R}_{\mathcal{A}}\left(\mathbf{w}(t)\right)\right]=0$. The derivation is tedious and the outline is given in the following. From (48) and (50), we have

$$\left(\mathbf{D}(t)\right)^T\mathcal{R}_{\mathcal{A}}\left(\mathbf{w}(t)\right)=\hat{x}(t)+\sum_{k\in\mathbb{Z}}\int_{t_k}^{t_{k+1}}\left(x(\tau)-\hat{x}(\tau)\right)d\tau\frac{\pi_k(t)}{t_{k+1}-t_k}$$

where

$$\hat{x}(t)=u^{\mathrm{I}}(t)\cos(2\pi f_0 t)-u^{\mathrm{Q}}(t)\sin(2\pi f_0 t)$$

Then

$$\begin{aligned}&\Re\left[x(t)-\left(\mathbf{D}(t)\right)^T\mathcal{R}_{\mathcal{A}}\left(\mathbf{w}(t)\right)\right]\\&=\Re\left(\left(x(t)-\hat{x}(t)\right)-\left(\sum_{k\in\mathbb{Z}}\int_{t_k}^{t_{k+1}}\left(x(\tau)-\hat{x}(\tau)\right)d\tau\frac{\pi_k(t)}{t_{k+1}-t_k}\right)\right)=\begin{bmatrix}①\\②\end{bmatrix}\end{aligned}$$

with

$$\begin{aligned}①=&\sum_{k\in\mathbb{Z}}\int_{t_k}^{t_{k+1}}\left[x(\tau)-\hat{x}(\tau)\right]d\tau\frac{\pi_k(t)}{t_{k+1}-t_k}\cos(2\pi f_0 t)-\\&\sum_{l\in\mathbb{Z}}\int_{t_\ell}^{t_{\ell+1}}\left[\sum_{k\in\mathbb{Z}}\int_{t_k}^{t_{k+1}}\left(x(\tau)-\hat{x}(\tau)\right)d\tau\frac{\pi_k(\eta)}{t_{k+1}-t_k}\right]d\eta\frac{\pi_\ell(t)}{t_{\ell+1}-t_\ell}\cos(2\pi f_0 t)\end{aligned}$$

$$\begin{aligned}②=&-\sum_{k\in\mathbb{Z}}\int_{t_k}^{t_{k+1}}\left[x(\tau)-\hat{x}(\tau)\right]d\tau\frac{\pi_k(t)}{t_{k+1}-t_k}\sin(2\pi f_0 t)+\\&\sum_{l\in\mathbb{Z}}\int_{t_\ell}^{t_{\ell+1}}\left[\sum_{k\in\mathbb{Z}}\int_{t_k}^{t_{k+1}}\left(x(\tau)-\hat{x}(\tau)\right)d\tau\frac{\pi_k(\eta)}{t_{k+1}-t_k}\right]d\eta\frac{\pi_\ell(t)}{t_{\ell+1}-t_\ell}\sin(2\pi f_0 t)\end{aligned}$$

Note that

$$\sum_{\ell\in\mathbb{Z}}\int_{t_\ell}^{t_{\ell+1}}\frac{\pi_k(\eta)}{t_{k+1}-t_k}d\eta=\begin{cases}1, & k=\ell\\0, & k\neq\ell\end{cases}$$

We have that $①=0$ and $②=0$. Thus $\mathcal{R}_{\mathcal{A}}\left(\mathcal{R}_{\mathcal{A}}\left(\mathbf{w}(t)\right)\right)=\mathcal{R}_{\mathcal{A}}\left(\mathbf{w}(t)\right)$.

Next let us show that $\mathcal{R}_{\mathcal{A}}$ has the range $\mathcal{A}$. By (47), for any $\mathrm{w}(t)=\left[u^{\mathrm{I}}(t),u^{\mathrm{Q}}(t)\right]^T\in\mathcal{A}$, we have

$$\int_{t_k}^{t_{k+1}}\left[u^{\mathrm{I}}(t)\cos(2\pi f_0 t)-u^{\mathrm{Q}}(t)\sin(2\pi f_0 t)\right]dt=\int_{t_k}^{t_{k+1}}x(u)du$$

Then

$$\begin{aligned}
\mathcal{R}_{\mathcal{A}}\left(\mathbf{w}(t)\right) &= \mathbf{w}(t) + \Re\left(x(t) - \left(\mathbf{D}(t)\right)^T \mathbf{w}(t)\right) \\
&= \mathbf{w}(t) + \Re\left(x(t) - \hat{x}(t)\right) \\
&= \mathbf{w}(t) + \begin{bmatrix} 2\sum_{k\in\mathbb{Z}}\int_{t_k}^{t_{k+1}}\left[x(\tau) - \hat{x}(\tau)\right]d\tau \dfrac{\pi_k(t)}{t_{k+1} - t_k}\cos\left(2\pi f_0 t\right) \\ -2\sum_{k\in\mathbb{Z}}\int_{t_k}^{t_{k+1}}\left[x(\tau) - \hat{x}(\tau)\right]d\tau \dfrac{\pi_k(t)}{t_{k+1} - t_k}\sin\left(2\pi f_0 t\right) \end{bmatrix}. \\
&= \mathbf{w}(t) + \begin{bmatrix} 0 \\ 0 \end{bmatrix} = \mathbf{w}(t)
\end{aligned}$$

Finally, we show that $\mathcal{R}_{\mathcal{A}}$ is firmly nonexpansive. From [29], it is sufficient to show that

$$\mathcal{R}_{\mathcal{A}} = \frac{1}{2}\mathcal{I}_{\mathcal{A}} + \frac{1}{2}\mathcal{N}_{\mathcal{A}}$$

where $\mathcal{I}_{\mathcal{A}}$ is an identity operator and $\mathcal{N}_{\mathcal{A}}$ is an nonexpansive operator. The above equation is equivalent to $\mathcal{N}_{\mathcal{A}} = 2\mathcal{R}_{\mathcal{A}} - \mathcal{I}_{\mathcal{A}}$ or

$$\begin{aligned}
\mathcal{N}_{\mathcal{A}}\left(\mathbf{w}(t)\right) &= 2\mathcal{R}_{\mathcal{A}}\left(\mathbf{w}(t)\right) - \mathcal{I}_{\mathcal{A}}\left(\mathbf{w}(t)\right) \\
&= 2\mathcal{I}_{\mathcal{A}}\left(\mathbf{w}(t)\right) + 2\Re\left(x(t) - \left(\mathbf{D}(t)\right)^T \mathbf{w}(t)\right) - \mathcal{I}_{\mathcal{A}}\left(\mathbf{w}(t)\right) \\
&= \mathcal{I}_{\mathcal{A}}\left(\mathbf{w}(t)\right) + 2\Re\left(x(t) - \left(\mathbf{D}(t)\right)^T \mathbf{w}(t)\right)
\end{aligned}$$

Then to show that $\mathcal{N}_{\mathcal{A}}$ is nonexpansive, it is enough to show $\left\|\mathcal{N}\left(\mathbf{w}_1(t)\right) - \mathcal{N}\left(\mathbf{w}_2(t)\right)\right\| \le \left\|\left(\mathbf{w}_1(t)\right) - \left(\mathbf{w}_2(t)\right)\right\|$ for any $\mathbf{w}_1(t)$ and $\mathbf{w}_2(t)$ in $\boldsymbol{L}^2(\mathbb{R})$. Note

$$\begin{aligned}
\left\|\mathcal{N}\left(\mathbf{w}_1(t)\right) - \mathcal{N}\left(\mathbf{w}_2(t)\right)\right\| &= \left\| \begin{matrix} \mathcal{I}_{\mathcal{A}}\left(\mathbf{w}_1(t)\right) + 2\Re\left(x(t) - \left(\mathbf{D}(t)\right)^T \mathbf{w}_1(t)\right) - \\ \left(\mathcal{I}_{\mathcal{A}}\left(\mathbf{w}_2(t)\right) + 2\Re\left(x(t) - \left(\mathbf{D}(t)\right)^T \mathbf{w}_2(t)\right)\right) \end{matrix} \right\| \\
&= \left\| \begin{matrix} \mathcal{I}_{\mathcal{A}}\left(\mathbf{w}_1(t)\right) - \mathcal{I}_{\mathcal{A}}\left(\mathbf{w}_1(t)\right) - 2\Re\left(\left(\mathbf{D}(t)\right)^T \mathbf{w}_1(t)\right) + \\ 2\Re\left(\left(\mathbf{D}(t)\right)^T \mathbf{w}_2(t)\right) \end{matrix} \right\| \\
&= \left\| \mathcal{I}_{\mathcal{A}}\left(\mathbf{w}_1(t) - \mathbf{w}_1(t)\right) - 2\Re\left(\left(\mathbf{D}(t)\right)^T \left(\mathbf{w}_1(t) - \mathbf{w}_2(t)\right)\right) \right\|
\end{aligned}$$

Let $\Re_{\mathrm{D}}\left(\mathbf{w}(t)\right) = \Re\left(\left(\mathbf{D}(t)\right)^T \left(\mathbf{w}(t)\right)\right)$. We have

$$\begin{aligned}\left\|\mathcal{N}\left(\mathbf{w}_1(t)\right)-\mathcal{N}\left(\mathbf{w}_2(t)\right)\right\| &= \left\|\mathcal{I}_{\mathcal{A}}\left(\mathbf{w}_1(t)-\mathbf{w}_1(t)\right)-2\mathfrak{R}_{\mathrm{D}}\left(\mathbf{w}_1(t)-\mathbf{w}_2(t)\right)\right\| \\ &= \left\|\left(\mathcal{I}_{\mathcal{A}}-2\mathfrak{R}_{\mathrm{D}}\right)\left(\mathbf{w}_1(t)-\mathbf{w}_1(t)\right)\right\| \\ &\le \left\|\mathcal{I}_{\mathcal{A}}-2\mathfrak{R}_{\mathrm{D}}\right\|\left\|\mathbf{w}_1(t)-\mathbf{w}_1(t)\right\|\end{aligned}$$

where $\left\|\mathcal{I}_{\mathcal{A}}-2\mathfrak{R}_{\mathrm{D}}\right\| = \sup_{\mathrm{w}}\left(\left\|\left(\mathcal{I}_{\mathcal{A}}-2\mathfrak{R}_{\mathrm{D}}\right)\mathbf{w}(t)\right\| / \left\|\mathbf{w}(t)\right\|\right)$. Now let us compute the norm.

$$\begin{aligned}\left\|\left(\mathcal{I}_{\mathcal{A}}-2\mathfrak{R}_{\mathrm{D}}\right)\mathbf{w}(t)\right\|^2 &= \left\|\begin{bmatrix} u^{\mathrm{I}}(t)-2\mathfrak{R}^{\mathrm{I}}\left(\hat{x}(t)\right) \\ u^{\mathrm{Q}}(t)-2\mathfrak{R}^{\mathrm{Q}}\left(\hat{x}(t)\right)\end{bmatrix}\right\|^2 \\ &= \int_{-\infty}^{\infty} \begin{matrix}\left(u^{\mathrm{I}}(\eta)-2\sum_{k\in\mathbb{Z}}\int_{t_k}^{t_{k+1}}\tilde{x}(\tau)d\tau\frac{\pi_k(\eta)\cos\left(2\pi f_0\eta\right)}{t_{k+1}-t_k}\right)^2 + \\ \left(u^{\mathrm{Q}}(\eta)+2\sum_{k\in\mathbb{Z}}\int_{t_k}^{t_{k+1}}\tilde{x}(\tau)d\tau\frac{\pi_k(\eta)\sin\left(2\pi f_0\eta\right)}{t_{k+1}-t_k}\right)^2\end{matrix}\, d\eta \\ &= \sum_{\ell\in\mathbb{Z}}\left(\int_{t_\ell}^{t_{\ell+1}}\left(\left(u^{\mathrm{I}}(\eta)\right)^2+\left(u^{\mathrm{Q}}(\eta)\right)^2\right)d\eta\right) - \\ &\quad 4\sum_{\ell\in\mathbb{Z}}\left(\int_{t_\ell}^{t_{\ell+1}}\hat{x}(\tau)d\tau\right)^2\frac{1}{t_{\ell+1}-t_\ell}+4\sum_{\ell\in\mathbb{Z}}\left(\int_{t_\ell}^{t_{\ell+1}}\hat{x}(\tau)d\tau\right)^2\frac{1}{t_{\ell+1}-t_\ell} \\ &= \sum_{\ell\in\mathbb{Z}}\left(\int_{t_\ell}^{t_{\ell+1}}\left(\left(u^{\mathrm{I}}(\eta)\right)^2+\left(u^{\mathrm{Q}}(\eta)\right)^2\right)d\eta\right) \\ &= \int_{-\infty}^{\infty}\left(u^{\mathrm{I}}(\eta)\right)^2+\left(u^{\mathrm{Q}}(\eta)\right)^2 d\eta \\ &= \left\|\mathbf{w}(t)\right\|^2\end{aligned}$$

Thus, $\left\|\mathcal{I}_{\mathcal{A}}-2\mathfrak{R}_{\mathrm{D}}\right\| = \sup_{\mathrm{w}}\left(\left\|\left(\mathcal{I}_{\mathcal{A}}-2\mathfrak{R}_{\mathrm{D}}\right)\mathbf{w}(t)\right\| / \left\|\mathbf{w}(t)\right\|\right) = 1$. We get

$$\begin{aligned}\left\|\mathcal{N}\left(\mathbf{w}_1(t)\right)-\mathcal{N}\left(\mathbf{w}_2(t)\right)\right\| &\le \left\|\mathcal{I}-2\mathfrak{R}_{\mathrm{D}}\right\|\left\|\mathbf{w}_1(t)-\mathbf{w}_1(t)\right\| \\ &\le \left\|\mathbf{w}_1(t)-\mathbf{w}_1(t)\right\|\end{aligned}.$$

□

*Proof of Lemma* 2: As in the proof of Lemma 1, we will prove that the operator is idempotent, has as the range the space $\mathcal{B}_{\mathrm{BL}}$ in $\boldsymbol{L}^2(\mathbb{R})$, and is firmly nonexpansive, respectively.

First, we show that $\mathcal{R}_{\mathcal{B}_{\mathrm{BL}}}$ is idempotent. Note that $g_{\mathrm{BL}}(t)$ is an ideal low-pass filter. Then $g_{\mathrm{BL}}(t) * g_{\mathrm{BL}}(t) = g_{\mathrm{BL}}(t)$. We have

$$\mathcal{R}_{\mathcal{B}_{\mathrm{BL}}}\left(\mathcal{R}_{\mathcal{B}_{\mathrm{BL}}}\left(\mathbf{w}(t)\right)\right) = \begin{bmatrix} u^{\mathrm{I}}(t) * g_{\mathrm{BL}}(t) * g_{\mathrm{BL}}(t) \\ u^{\mathrm{Q}}(t) * g_{\mathrm{BL}}(t) * g_{\mathrm{BL}}(t)\end{bmatrix} = \begin{bmatrix} u^{\mathrm{I}}(t) * g_{\mathrm{BL}}(t) \\ u^{\mathrm{Q}}(t) * g_{\mathrm{BL}}(t)\end{bmatrix} = \mathcal{R}_{\mathcal{B}_{\mathrm{BL}}}\left(\mathbf{w}(t)\right)$$

We now show that $\mathcal{R}_{\mathcal{B}_{\mathrm{BL}}}\left(\mathbf{w}(t)\right)$ has as range the space $\mathcal{B}_{\mathrm{BL}}$ in $\boldsymbol{L}^2(\mathbb{R})$. Note that $\mathbf{w}(t)=\left[u^{\mathrm{I}}(t),u^{\mathrm{Q}}(t)\right]^T$. Then

$$\mathbf{w}(t)*g(t)=\begin{bmatrix}u^{\mathrm{I}}(t)\\u^{\mathrm{Q}}(t)\end{bmatrix}*g(t)=\begin{bmatrix}u^{\mathrm{I}}(t)*g_{\mathrm{BL}}(t)\\u^{\mathrm{Q}}(t)*g_{\mathrm{BL}}(t)\end{bmatrix}=\begin{bmatrix}u^{\mathrm{I}}(t)\\u^{\mathrm{Q}}(t)\end{bmatrix}=\mathbf{w}(t)$$

Thus $\mathbf{w}(t)*g_{\mathrm{BL}}(t)\in\mathcal{B}_{\mathrm{BL}}$.

Finally, $\mathcal{R}_{\mathcal{B}_{\mathrm{BL}}}$ is firmly non-expansive, since $\mathcal{R}_{\mathcal{B}_{\mathrm{BL}}}\left(\mathrm{w}(t)\right)=\left[u^{\mathrm{I}}(t),u^{\mathrm{Q}}(t)\right]^T*g_{\mathrm{BL}}(t)$ is an orthogonal projection operator.

□

*Proof of Lemma* 3: Let $\mathbf{w}_1(t)\in\mathcal{A}$ and $\mathbf{w}_2(t)\in\mathcal{A}$. For any convex combination of $\mathbf{w}_1(t)$ and $\mathbf{w}_2(t)$, we have $\mathbf{w}_3(t)=\lambda\mathbf{w}_1(t)+(1-\lambda)\mathbf{w}_2(t)$ with $\lambda\in[0,1]$. Then if $\mathbf{w}_3(t)\in\mathcal{A}$, $\mathcal{A}$ is convex. Let $\mathrm{w}_1(t)=\left[u_1^{\mathrm{I}}(t),u_1^{\mathrm{Q}}(t)\right]^T$, $\mathrm{w}_2(t)=\left[u_2^{\mathrm{I}}(t),u_2^{\mathrm{Q}}(t)\right]^T$, and $\mathrm{w}_3(t)=\left[u_3^{\mathrm{I}}(t),u_3^{\mathrm{Q}}(t)\right]^T$. We have $u_3^{\mathrm{I}}(t)=\lambda u_1^{\mathrm{I}}(t)+(1-\lambda)u_2^{\mathrm{I}}(t)$ and $u_3^{\mathrm{Q}}=\lambda u_1^{\mathrm{Q}}(t)+(1-\lambda)u_2^{\mathrm{Q}}(t)$. From (47), we can get

$$\begin{aligned}&\left\langle f_k^{\mathrm{I}}(t),u_3^{\mathrm{I}}(t)\right\rangle-\left\langle f_k^{\mathrm{Q}}(t),u_3^{\mathrm{Q}}(t)\right\rangle\\&=\left\langle f_k^{\mathrm{I}}(t),\lambda u_1^{\mathrm{I}}(t)+(1-\lambda)u_2^{\mathrm{I}}(t)\right\rangle-\left\langle f_k^{\mathrm{Q}}(t),\lambda u_1^{\mathrm{Q}}(t)+(1-\lambda)u_2^{\mathrm{Q}}(t)\right\rangle\\&=\lambda\left[\left\langle f_k^{\mathrm{I}}(t),u_1^{\mathrm{I}}(t)\right\rangle-\left\langle f_k^{\mathrm{Q}}(t),u_2^{\mathrm{Q}}(t)\right\rangle\right]+(1-\lambda)\left[\left\langle f_k^{\mathrm{I}}(t),u_2^{\mathrm{I}}(t)\right\rangle-\left\langle f_k^{\mathrm{Q}}(t),u_2^{\mathrm{Q}}(t)\right\rangle\right]\\&=\lambda y_k+(1-\lambda)y_k\\&=y_k,k\in\mathbb{Z}\end{aligned}$$

Then $\mathbf{w}_3(t)\in\mathcal{A}$.

□

*Proof of Lemma* 4: From the proof of Lemma 3, we know that the Fourier transform of $u_3^{\mathrm{I}}(t)$ is the convex combination of the Fourier transforms $u_1^{\mathrm{I}}(t)$ and $u_2^{\mathrm{I}}(t)$. Similarly, do $u_3^{\mathrm{Q}}(t)$. Then the Fourier transforms of $u_3^{\mathrm{I}}(t)$ and $u_3^{\mathrm{Q}}(t)$ have the same support as those of $u_2^{\mathrm{I}}(t)$, $u_2^{\mathrm{Q}}(t)$, $u_1^{\mathrm{I}}(t)$ and $u_1^{\mathrm{Q}}(t)$. Therefore $\mathrm{w}_3(t)\in\mathcal{B}_{\mathrm{BL}}$.

□